\documentclass[intlimits,twoside,a4paper]{article}

\usepackage[utf8]{inputenc}

\usepackage[eqsecnum]{cmpj3}

\issue{2023}{26}{3}{33602}
\doinumber{10.5488/CMP.26.33602}
\title[Proportional correlation between heat capacity and thermal expansion]%
{Proportional correlation between heat capacity and thermal expansion of atomic, molecular crystals and carbon nanostructures%
}

\author[M. S. Barabashko, A. I. Krivchikov, R. Basnukaeva, O.~A.~Korolyuk,  A. Je\.zowski]{
M. S. Barabashko\orcid{0000-0003-3168-7119}\refaddr{label1}\thanks{Corresponding author: \email{barabaschko@ilt.kharkov.ua}.}, 
 A. I. Krivchikov\orcid{0000-0001-5375-439X}\refaddr{label1,label2},
  R. Basnukaeva\orcid{0000-0003-4706-3837}\refaddr{label1},
  O. A. Korolyuk\orcid{0000-0003-4282-297X}\refaddr{label1, label3},
   A. Je\.zowski\orcid{0000-0002-4955-8051}\refaddr{label2}
 }
\addresses{
\addr{label1} B.Verkin Institute for Low Temperature Physics and Engineering of the National Academy of Sciences of Ukraine, 47 Nauka Ave., 61103 Kharkov, Ukraine
\addr{label2} Institute of Low Temperature and Structure
Research, Polish Academy of Sciences, 50-422 Wroclaw
\addr{label3} Donostia International Physics Center (DIPC), Paseo Manuel de Lardizabal 4, 20018 Donostia, San Sebastian, Spain
}


\date{Received January 31, 2023, in final form April 03, 2023}

\begin{document}

\maketitle

\begin{abstract}
Correlation between thermal expansions $\beta(T)$ and heat capacity $C(T)$  of atomic and molecular crystals, amorphous materials with a structural disorder, carbon nanomaterials (fullerite C$_{60}$,  bundles SWCNTs of single-walled carbon nanotubes) was analyzed. The influence of the contribution to the coefficient of linear thermal expansion $\alpha_\textrm{Xe}(T)$ of Xe atoms adsorbed on the SWCNTs bundles is considered. The proportional correlation  was found between the contribution to the coefficient of linear thermal expansion $\alpha_\textrm{Xe}(T)$  and the normalized to the gas constant heat capacity $C_\textrm{Xe}(T)/R$ of Xe atoms adsorbed on the  SWCNTs bundles. The proportional correlation $(\beta/\beta^*)  \sim (C_\textrm{V}/R)$ with the parameter $\beta^*$ for the bulk thermal expansion coefficient for cryocrystals is proposed.  In the case of atomic crystals such as Xe and Ar, the proportional correlation  $(\beta/\beta^*)  \sim (C_\textrm{V}/R)$ is observed in the temperature range from the lowest experimental to temperatures where  $C_\textrm{V}/R  \approx 2.3$. The correlation is not observed in the temperatures where $2.3<C_V/R<3$ (classical Dulong-Petit law). It was found that the universal proportional correlation is also observed for molecular crystals with linear symmetry, such as CO$_{2}$, CO, and N$_{2}$O if the normalized heat capacity  below the values $C_\textrm{V}/R \approx 3 \div 3.5$. It indicates that the proportional correlation between thermal expansions $(\beta/\beta^*)$ and heat capacity $(C_\textrm{V}/R)$ is related not only to the translational, but also to the rotational degrees of freedom of the molecule in the crystal. In the case of the C$_{60}$, molecular crystal with translational and rotational degrees of freedom and intramolecular vibrations, the discussed above correlation occurs below the values of normalized heat capacity $C_\textrm{V}/R \approx 7.5$. In strongly anisotropic systems, such as systems of compacted bundles of single-walled carbon nanotubes and SWCNTs bundles with adsorbed Xe atoms, this universal dependence appears in a limited temperature range that does not include the lowest temperatures. A qualitative explanation of the observed correlation is proposed.
%
%
\printkeywords
%
\end{abstract}

\section{Introduction}


The heat capacity $C(T)$ and the volumetric thermal expansion coefficient $\alpha(T)$ are fundamental and important thermodynamic parameters for solids that are widely studied using many theoretical and experimental methods. Studies of heat capacity and thermal expansion are powerful tools for understanding lattice vibrations, phase transitions, tunneling, and quantum effects in ordered and disordered systems~\cite{white1993solids}. Knowledge of low-temperature thermal properties of condensed matter materials, including crystals~\cite{gavrilko1999structure}, amorphous materials~\cite{hassaine2012low}, glasses~\cite{manzhelii1998thermodynamic} and nanomaterials~\cite{barabashko2021calorimetric, bagatskii2021size,barabashko2017low} are important both for the development of electric and photonic nanodevices, energy storage/conversion and thermal control devices~\cite{dolbin2019thermal, bagatskii2014low}, as well as for the creation of new composite functional materials~\cite{vinnikov2022analysis, dolbin2020influence}, etc.

    At low temperatures, the character of $C(T)$ and $\beta(T)$ depends on the quantum nature of atomic vibrations, the dimensions of nanosystems, structural features, and the presence of impurities~\cite{krivchikov2022role, rusakova2020possible, manzhelii2017influence, bagatskii2017heat, dolbin2017thermal}. The experimentally measured heat capacity at constant pressure $C_\textrm{P}(T)$ and the theoretical phonon heat capacity for constant volume $C_\textrm{V}(T)$ are close at low temperatures. The  temperature increase leads to a difference between $C_\textrm{P}(T)$ and $C_\textrm{V}(T)$ due to thermal expansion and other factors such as vacancy formation, molecular rotation and melting processes~\cite{bagatskii2016heat, manzhelii1971thermal, barron2012heat, ramos2013low}. To determine the difference between $C_\textrm{P}(T) - C_\textrm{V}(T)$ both $\beta(T)$  and the parameters $\chi(T)$  (isotermal compression coefficient), $V(T)$ (molar volume) and $\gamma(T)$ (Gr\"uneisen coefficient) are required~\cite{gruneisen1912theorie}. Some of these parameters may be missing in the literature for the new nanomaterials.

The proportional correlation between the linear coefficient of thermal expansion $\alpha(T)$ and $C_\textrm{V}(T)$ is proposed theoretically for 1D atomic chains~\cite{pathak2022thermal, meincke1962two, leibfried1961theory}. A proportionality between the experimental data $C_P(T)$ and $\beta(T)$ is observed for 3D materials, as noted by Tang et al.~\cite{tang2021scaling}. Their analysis for some standard solids such as inert monatomic solids (Xe, Kr and Ar), metal oxides (MgO and Al$_\textrm{2}$O$_\textrm{3}$) and metals (W, Ta, Mo, Pt, Be, Cu, Al and Hg) is based on the previous works of Bodryakov~\cite{bodryakov2014correlation, bodryakov2015correlation}. It is possible to suggest that the proportional dependence between $C_\textrm{P}(T)$ and $\beta(T)$ can be generalized for all solids in a wide range of temperatures from a few Kelvin to the melting temperature. However, as far as we know, previous changes in the nature of temperature dependencies of heat capacity and thermal expansion of molecular cryocrystals, for which the influence of rotational degrees of freedom is essential, were not considered in the frames of such approach~\cite{gavrilko1999structure, manzheliui1997physics, venables1977rare}. Moreover, carbon nanomaterials that have the  manifestation of the features of their unique 1D and 2D geometry on temperature dependencies $C_\textrm{P}(T)$ and $\beta(T)$ as well the fundamentally different energy spectrum compared to classical 3D materials, is not considered within the frames of the approach discussed in references~\cite{tang2021scaling,bodryakov2014correlation, bodryakov2015correlation}.

Recently, experimental results of $C(T)$ and the linear coefficient of thermal expansion $\alpha(T)$ for bulk samples prepared by mechanical compressing the powder of SWCNTs bundles were compared by Bagatskii et al.~\cite{bagatskii2012specific}. Bundles of single-walled carbon nanotubes are highly anisotropic systems. The low-temperature thermal properties of SWCNTs bundles have a quasi-one-dimensional nature (the quasi-1D phonon spectrum and features in temperature dependencies of  $C(T)$ and $\alpha(T)$~\cite{bagatskii2012specific, schelling2003thermal}). It was found that the Gr\"uneisen parameter $\gamma(T) $ of SWCNTs bundles has a feature near 36 K. Above 36~K, the Gr\"uneisen coefficient is  $\gamma(T) \approx 4$ and monotonously decreases with the decreasing of  temperature below 36~K~\cite{bagatskii2012specific}. For orientational glass of fullerite  C$_\textrm{60}$, the contribution of tunneling levels and intramolecular vibrations to the heat capacity taken into account~\cite{barabashko2017low, bagatskii2015low}. The Gr\"uneisen coefficient of fullerite  C$_{60}$ is  $\gamma(T) \approx 3$, and the temperature dependences of $\alpha(T)$ and $C_\textrm{P}(T)$ are similar in the range from 5 to 63~K~\cite{bagatskii2015low}.

The aim of this study is to analyze the data of heat capacity and thermal expansion of crystalline and amorphous materials in order to determine the influence of the manifestation of the structure of solids on the proportional dependence between $C(T)$ and $\beta(T)$. We considered the systems of xenon atoms adsorbed on the surface of carbon nanotubes, atomic (Xe, Ar) and molecular cryocrystals (N$_{2}$, CO$_{2}$, CO, N$_{2}$O) and carbon nanostructures with a complex phonon spectrum. The ratio between thermal expansion and heat capacity for simple atomic crystals (only 3 translational degrees of freedom), molecular crystals (additional degrees of freedom associated with the rotational motion of the molecule) and carbon nanomaterials (strongly anisotropic systems, intramolecular vibrations and tunnel levels) were analyzed. The translational degrees of freedom of a molecule are associated with three independent directions of translational motion of the centre of mass of the molecule in three-dimensional space. Rotational degrees of freedom --- with independent axes of rotation of the molecules. The contribution of rotational degrees to the thermal properties of  crystals may be understood when considering the dynamics of a crystal consisting of a system of interacting oscillators generated by both translational and orientational degrees of freedom.
\newpage

\section{Experiment and discussion}
\subsection{Heat capacity of the Xe atoms adsorbed by SWCNTs bundles}
\begin{figure}[!htb]
	\centering
	\includegraphics[width=0.9\linewidth]{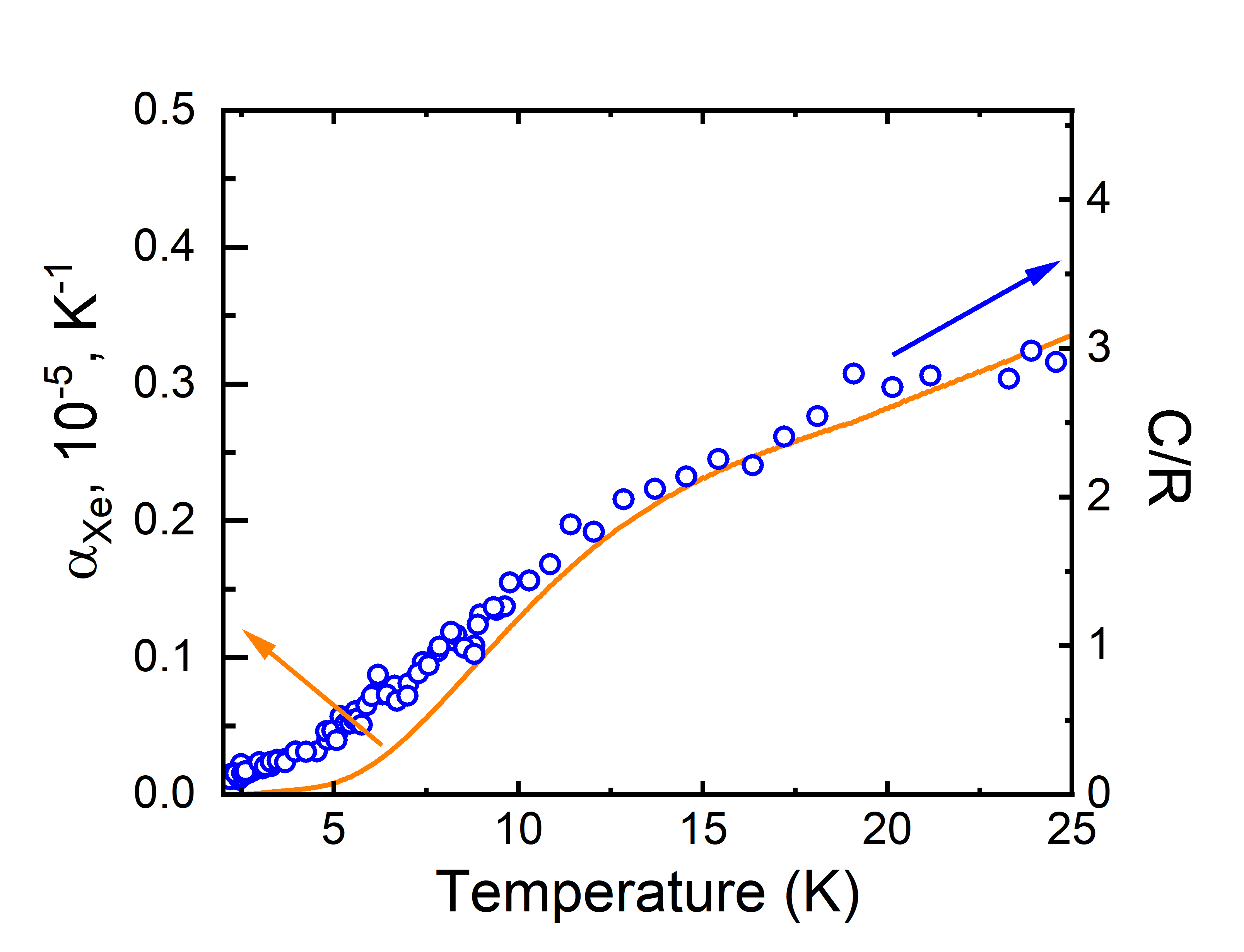}  
	\caption{(Colour online) $\alpha_\textrm{Xe}(T)$ vs. $C_\textrm{Xe}(T)/R$. Left-hand Y axis is thermal expansion $\alpha_\textrm{Xe}(T)$ of the Xe contribution to the $\alpha(T)$ systems of carbon nanotubes with adsorbed Xe atoms (solid line)~\cite{dolbin2009radial}.  Right-hand Y axis is heat capacity normalized to the gas constant $C(T)/R$: $C_\textrm{Xe}(T)$ is the Xe contribution to the heat capacity of the SWCNT$-$Xe system (experiment --- empty circles~\cite{bagatskii2013experimental}).}
	\label{fig:1}
\end{figure}

Figure~\ref{fig:1} shows the contributions of adsorbed Xe atoms to the heat capacity $C_\textrm{Xe}(T)$ (circles) and the thermal expansion $\alpha_\textrm{Xe}(T)$ (line) of the SWCNTs$-$Xe system. Heat capacity~\cite{bagatskii2013experimental, barabashko2015heat, bagatskii2014thermal, barabashko2021experimental}  and thermal expansion~\cite{dolbin2009radial, dolbin2008vb} of single-walled carbon nanotube (SWCNTs) bundles with adsorbed Xe atoms were measured for the same SWCNTs powder material mechanically compressed under pressure 1~GPa. The experimental heat capacity $C_\textrm{Xe}(T)$ for the contribution of the Xe adsorbate~\cite{bagatskii2013experimental}  agrees with the theoretical phonon heat capacity for 1D Xe chains at constant volume $C_\textrm{V}(T)$~\cite{manzhelii2015phonon}. 

\begin{figure}[!htb]
	\centering
	\includegraphics[width=0.9\linewidth]{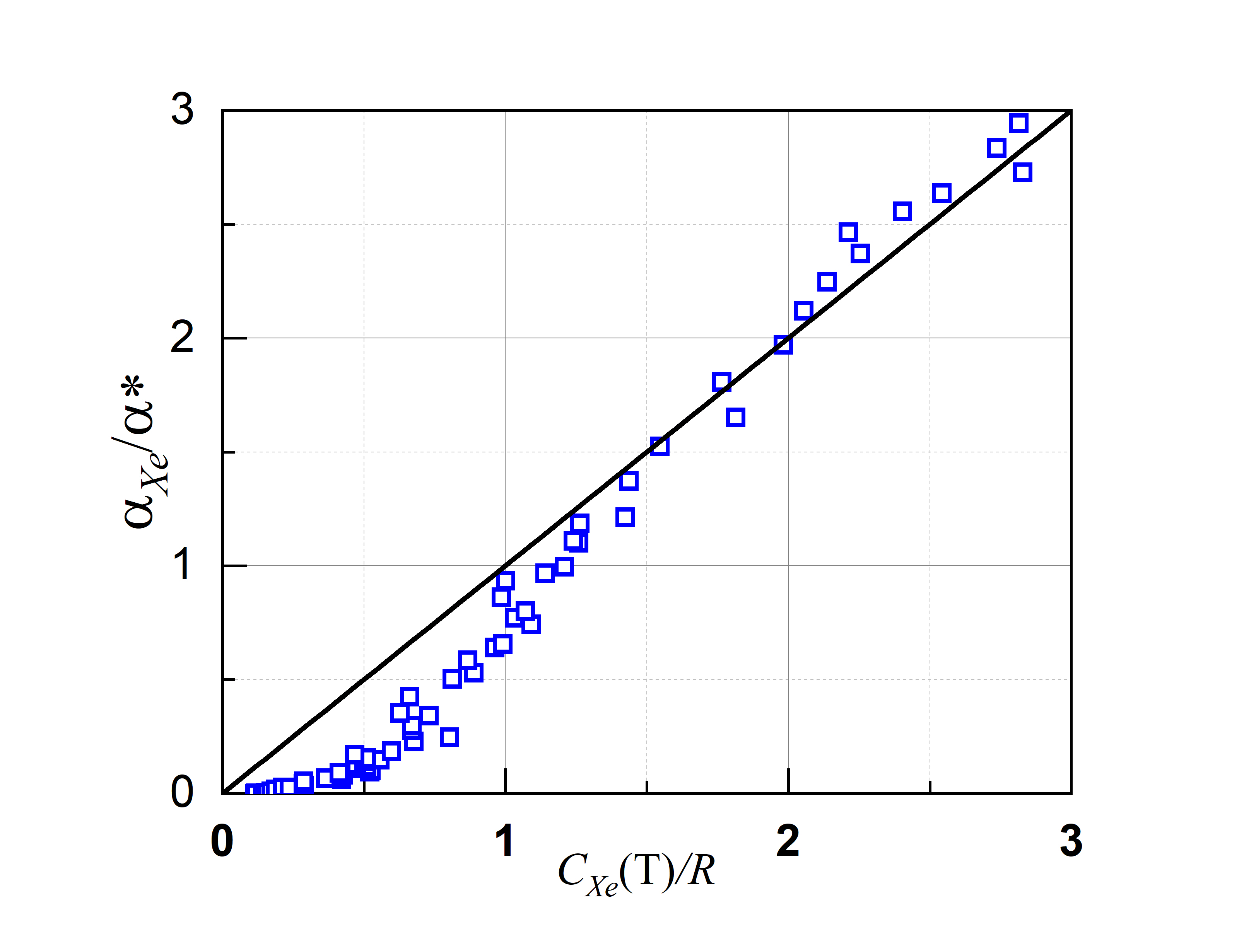}  
	\caption{(Colour online) Dependence $\alpha_\textrm{Xe}(T)/\alpha^*$ vs. $C_\textrm{Xe}(T)/R$ in normal scale  for  contributions of Xe atoms (empty circles) adsorbed by SWCNT bundles. Straight line shows the extrapolation of the experimental data by functions $\alpha_\textrm{Xe} \approx \alpha^*C_\textrm{Xe}(T)/R$, where $\alpha^*$ is a constant for the $C_\textrm{Xe}(T)/R > 1$.}
	\label{fig:2}
\end{figure}

In figure~\ref{fig:1}, a similar temperature dependence for  $\alpha_\textrm{Xe}(T)$ and $C_\textrm{Xe}(T)$ is observed in the temperature range from 10 to 30 K. It indicates that the influence of adsorbed xenon atoms makes the main positive contribution to the thermal expansion of the SWCNT--Xe systems. Near 10 K, the slope of the tangent to $\alpha_\textrm{Xe}(T)$ increases with decreasing temperature. Let us perform normalization by $\alpha^* = 10^{-6}$ K$^{-1}$. The coefficient $\alpha^*$ is chosen so that the relationship between $\alpha_\textrm{Xe}(T)/\alpha^*$ and $C_\textrm{Xe}(T)$ fits by the linear function $y = kx$, with the angular coefficient $k=1$, as shown in figure~\ref{fig:2}. In the case of the heat capacity $C_\textrm{Xe}(T)/R >1$, a typical for simple molecular crystals proportional dependence between  $\alpha_{Xe}(T)/\alpha^*$  and $C_{Xe}(T)/R$ is observed. For the heat capacity $C_\textrm{Xe}(T)/R <1$, there is a significant deviation from this dependence, which is associated with the anisotropy of the system of adsorbed atoms. The temperature dependence between $\alpha_\textrm{Xe}(T)/\alpha^*$  and $C_\textrm{Xe}(T)/R$ can be explained by the decrease of the Gr\"uneisen coefficient for the low-dimensional system of Xe adsorbates with the decrease of temperature below 10~K. Note that the total heat capacity of the SWCNT is less than $3R$~\cite{hone2000quantized}.

\subsection{Relation between volume thermal expansion $\beta$ and $C_\textrm{V}$ based on phenomenological theory}

According to the phenomenological theory of the thermal expansion of solids~\cite{landau2013statistical, gruneisen1908zusammenhang}, in considering thermal expansion from a thermodynamic point of view, the volume coefficient of thermal expansion depends on the entropy $S$:
\begin{equation}\label{1} 
 \beta = \chi_\textrm{T} \left(\frac{\partial S}{\partial V}\right)_T,
\end{equation}
where, $\chi_\textrm{T}$ is coefficient of isothermal compression. Heat capacity is also a function of entropy:
\begin{equation}\label{2}
 C_V = \frac{1}{T} \left(\frac{\partial S}{\partial T}\right)_\textrm{V}.
\end{equation}
Thus, there is a proportional relationship between the thermal expansion and the heat capacity of solids with the coefficient $\gamma$:
\begin{equation}\label{3} 
 \beta = \frac{\gamma \chi_\textrm{T} C_\textrm{V}}{V},
\end{equation}
where  $\gamma = -\left({\partial \ln\,T}/{\partial \ln\,V}\right)_\textrm{S}$ is Gr\"uneisen coefficient, which characterizes the change of temperature during an adiabatic change of volume~\cite{gruneisen1912theorie, gruneisen1908zusammenhang}. In the case when $\gamma$ does not depend on temperature, in the frames of the phenomenological theory of thermal expansion of solids between $\beta (T)$ and $C_\textrm{V}(T)/R$ it is possible to propose a constant $\beta^*$:
\begin{equation}\label{beta*} 
 \beta^* = \frac{\gamma \chi_\textrm{T} R}{ V} = \frac{3\alpha R}{C_\textrm{V}} .
\end{equation}
This constant $\beta^*$ can be found directly from a comparison of the temperature dependences of heat capacity and thermal expansion of new nanomaterials, without involving the missing data about the temperature dependences $\chi_\textrm{T}$, $V(T)$, $\gamma(T)$. Thus,   $\beta^*(T)$  is the normalization parameter for the bulk thermal expansion coefficient, which is the value of bulk thermal expansion coefficient of a solid while the heat capacity $C_\textrm{V}(T) = R$. Note, that using $\beta^*(T)$ we simplify the analysis of thermodynamic properties of low-dimensional nanomaterials such as carbon nanostructures. It was shown by Bagatskii et al. ~\cite{bagatskii2012specific} that the negative values of $\gamma(T)$ are observed for SWCNTs bundles with the decrease of temperature, wich is much different from the constant value $\gamma(T) \approx$ const for higher temperatures. Using $\beta^*(T)$  can also be useful for qualitative analysis of  $C_\textrm{P}(T)$ and $\gamma(T)$ of nanomaterials of the same nature or for determination of the temperature limits of the domination of the positive phonon contribution to the thermal properties of nanomaterials with high anisotropy or the absence of long-range structural order.

Within the framework of the dynamics of the crystal lattice, the parameter $\gamma$ can be obtained by the next equation with data  of heat capacity at a constant volume  $C_{\textrm{V},i}$, corresponding to one phonon with the $i$ index as:
\begin{equation}\label{4} 
 \gamma = \frac{\sum_{i=1} \gamma_i C_{\textrm{V},i}}{ \sum_{i=1} C_{\textrm{V},i}}.
\end{equation}
Here, the coefficient related to the change of the frequency of the phonon $\gamma_i$ at the volume changes is:
\begin{equation}\label{5} 
 \gamma_{i} = -\frac{\partial (\ln\omega_{i})}{\partial (\ln V)}.
\end{equation}
The summation in the equation~({\ref{4}}) is performed over all phonons in the first Brillouin zone. The Gr\"uneisen parameter is calculated by using the density of vibrational states $g(\omega)$  as follows:
\begin{equation}\label{6} 
\gamma = \frac{\int_{0}^{\omega_\textrm{max}} \gamma(\omega)\, g(\omega)\, C_V(\omega, T) \, \rd\omega}{\int_{0}^{\omega_\textrm{max}} g(\omega)\, C_V(\omega, T) \, \rd\omega}.
\end{equation}
In the Debay model, all frequencies are proportional to the Debay frequency 
$\omega_\textrm{D}$, the magnitude of which depends on the molar volume. Therefore, the Gr\"uneisen 
parameter becomes the same for all modes:
\begin{equation}\label{7} 
\gamma = -\frac{\partial (\ln\omega_\textrm{D})}{\partial (\ln V)}.
\end{equation}
The theoretically predicted linear correlation between the volume coefficient of thermal expansion and the thermal heat capacity was investigated for highly symmetrical atomic crystals by Garai~\cite{garai2006correlation}.

\subsection{Ratio between $\beta$ and the heat capacity $C_\textrm{V}$ of carbon nanomaterials}

\begin{figure}[!htb]
	\centering
	\includegraphics[width=0.45\linewidth]{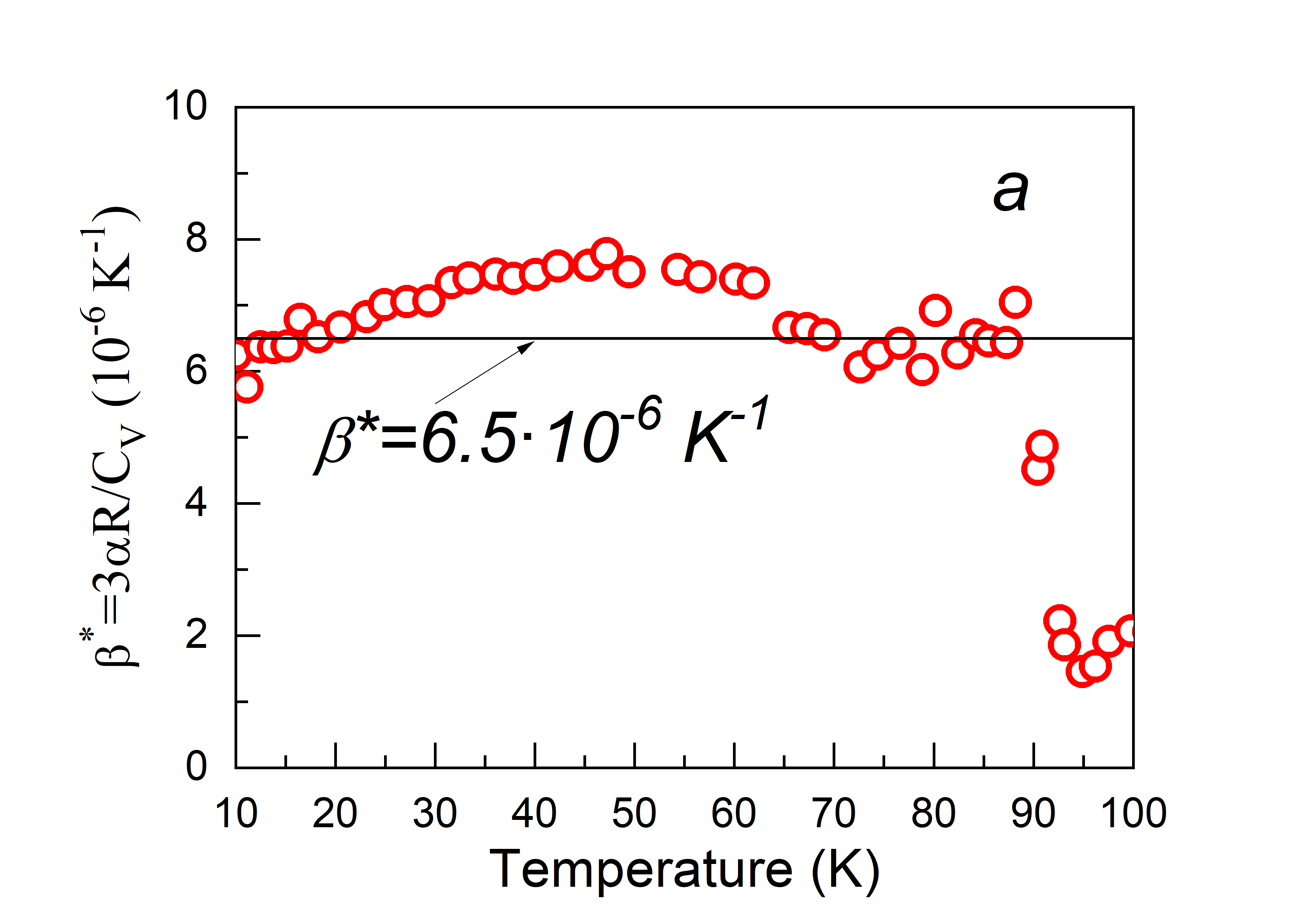}  
	\includegraphics[width=0.45\linewidth]{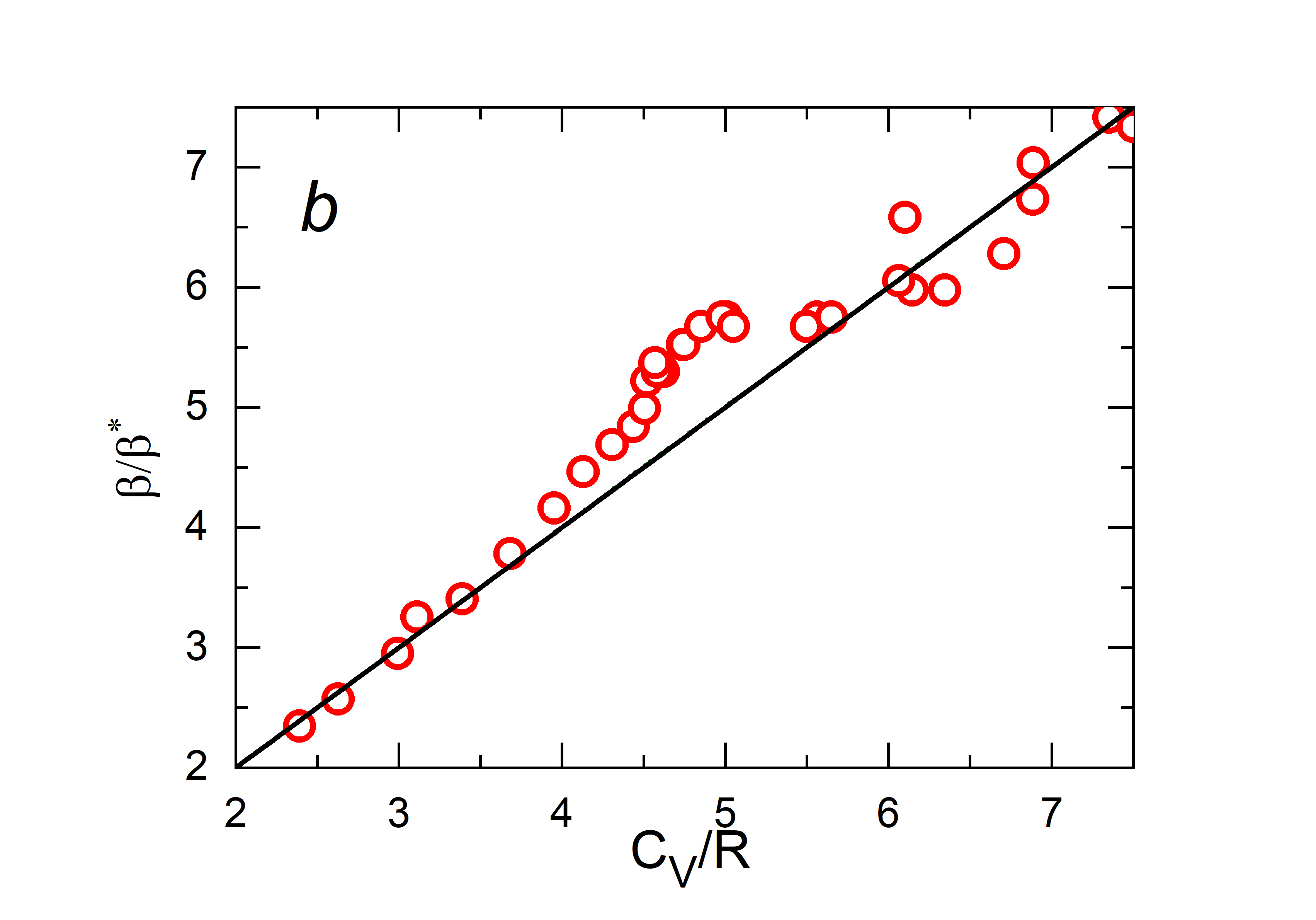}  
	\caption{ (Colour online)  (a) The calculated temperature dependence for   C$_{60}$ fullerite (red circles). Straight line is an approximation by the constant $\beta^* = 6.5\times10^{-6}$~K$^{-1}$. (b) The dependence of $\beta /  \beta^*$ on $C_\textrm{V}(T)/R$, plotted by using $\beta^* = 6.5\times10^{-6}$~K$^{-1}$, and the experimental data~\cite {barabashko2017low, bagatskii2015low,aleksandrovskii2005polyamorphism, gugenberger1992glass}; the line shows the proportional dependence.    }
	\label{fig:3}
\end{figure}

Let us look at the relationship between the experimental dependencies of  $C_\textrm{P}(T)$ and $\beta(T)$ carbon nanomaterials such as SWCNT bundles and fullerite C$_{60}$. The heat capacity of fullerite C$_{60}$  with a purity of 99.99\% at constant pressure was studied using an adiabatic calorimeter in the temperature range from 1.2 to 120~K~\cite {barabashko2017low, bagatskii2015low}. The difference $C_\textrm{P}(T) - C_\textrm{V}(T)$ is negligible below 120 K (less than 0.1\%)~\cite {aksenova1999analysis}. At the same time, the contribution of the 
    intramolecular vibrations to the heat capacity is significant above 50~K~\cite {bagatskii2015low}. Linear thermal expansion $\alpha(T)$  of polycrystalline C$_\textrm{60}$   was measured on the same C$_{60}$ material as the heat capacity below 20~K~\cite {aleksandrovskii2005polyamorphism}. Above 20 K, the bulk thermal expansion data were calculated as $\beta \approx 3\alpha$ using $\alpha$ data for monocrystal C$_\textrm{60}$~\cite {gugenberger1992glass}. The temperature dependence of $\beta^*$ [based on equation~(\ref{beta*})] in the case of 
    C$_{60}$  fullerite was calculated using data on heat capacity~\cite {barabashko2017low, bagatskii2015low} and thermal expansion~\cite {aleksandrovskii2005polyamorphism, gugenberger1992glass}. The obtained temperature dependence $\beta^* = 6.5\times 10^{-6}$~K$^{-1}$   for C$_{60}$  fullerite is shown in figure~\ref{fig:3}a. It can be seen that in a wide temperature range from 10 to 90 K, this dependence can be approximated by constant  $\beta^* = 6.5\times10^{-6}$~K$^{-1}$. The normalization of the thermal expansion of fullerite C$_\textrm{60}$ to the parameter $\beta^* = 6.5\times10^{-6}$~K$^{-1}$ was carried out the same way as in the case of the system of the single-walled carbon nanotubes with adsorbed Xe (SWCNTs-Xe). The obtained dependence  $\beta /  \beta^*$ vs. $C_\textrm{V}/R$ is shown in figure~\ref{fig:3}b. It can be seen that there is a direct proportional dependence in a wide temperature range between the volume thermal expansion coefficient and the heat capacity (line in figure~\ref{fig:3}b). This indicates a high agreement between the experimental data of $C_\textrm{P}(T)$ and $\beta(T)$ of C$_{60}$  fullerite. 

\begin{figure}[!htb]
	\centering
	\includegraphics[width=0.3\linewidth]{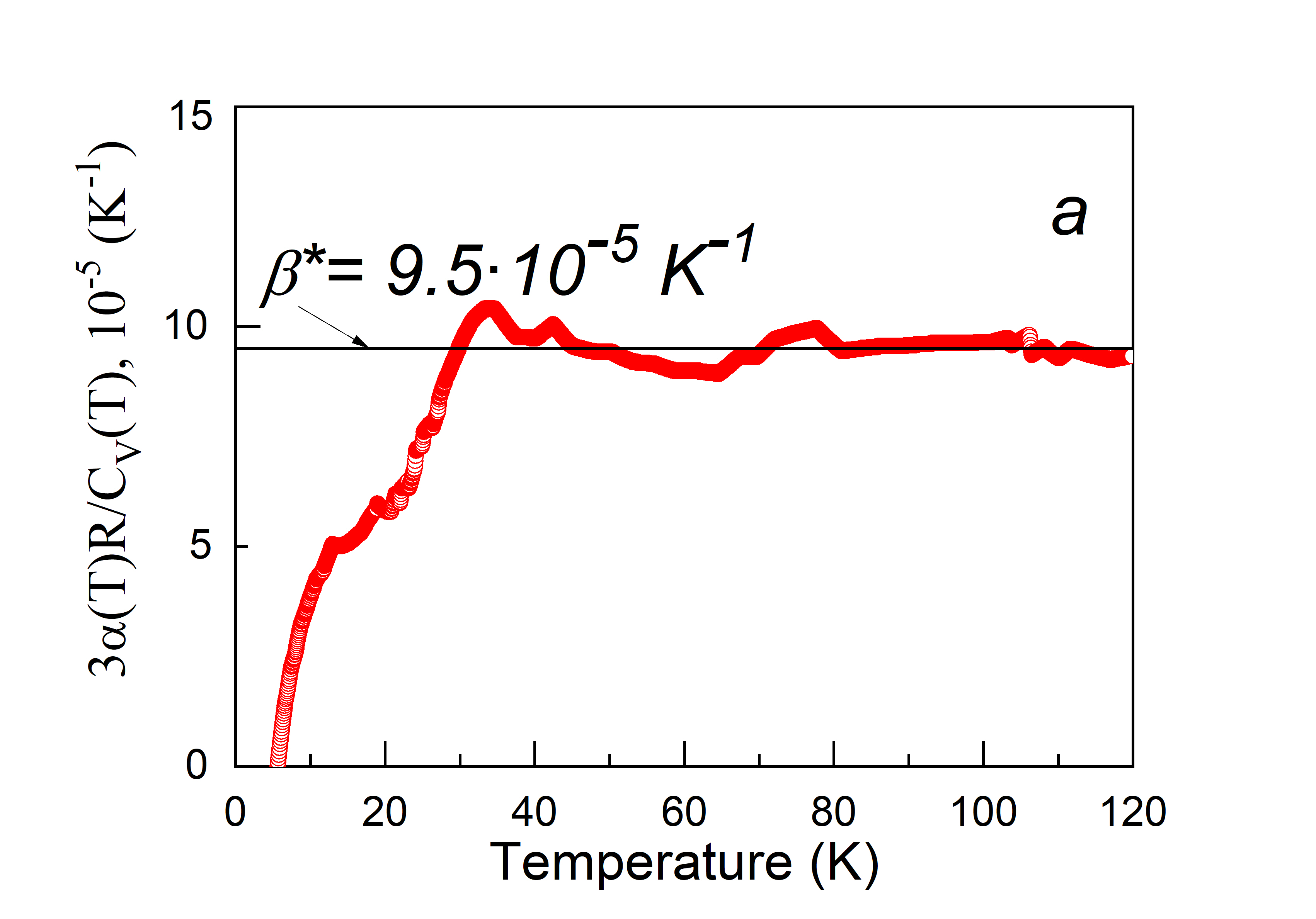}  
	\includegraphics[width=0.3\linewidth]{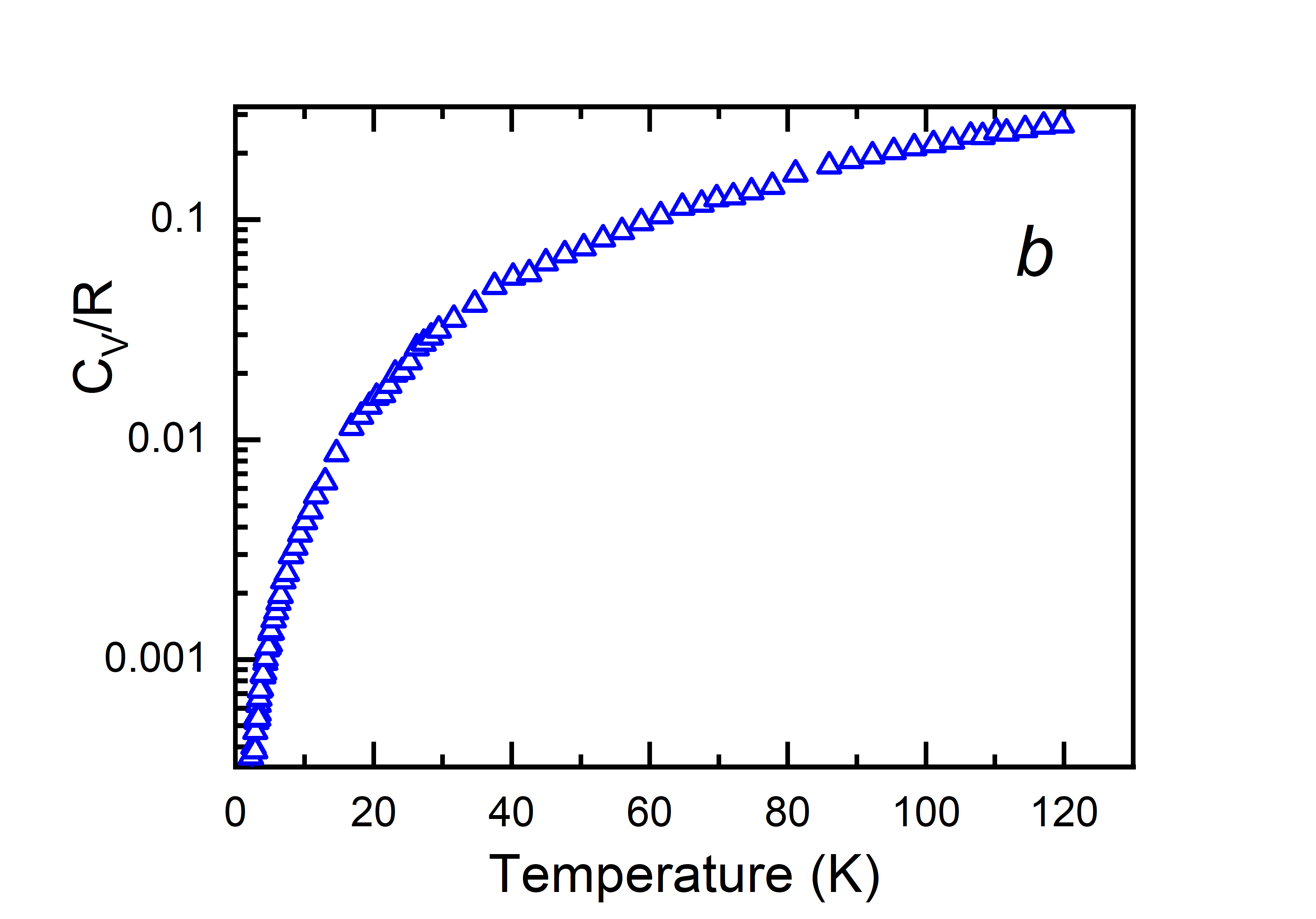}  
	\includegraphics[width=0.3\linewidth]{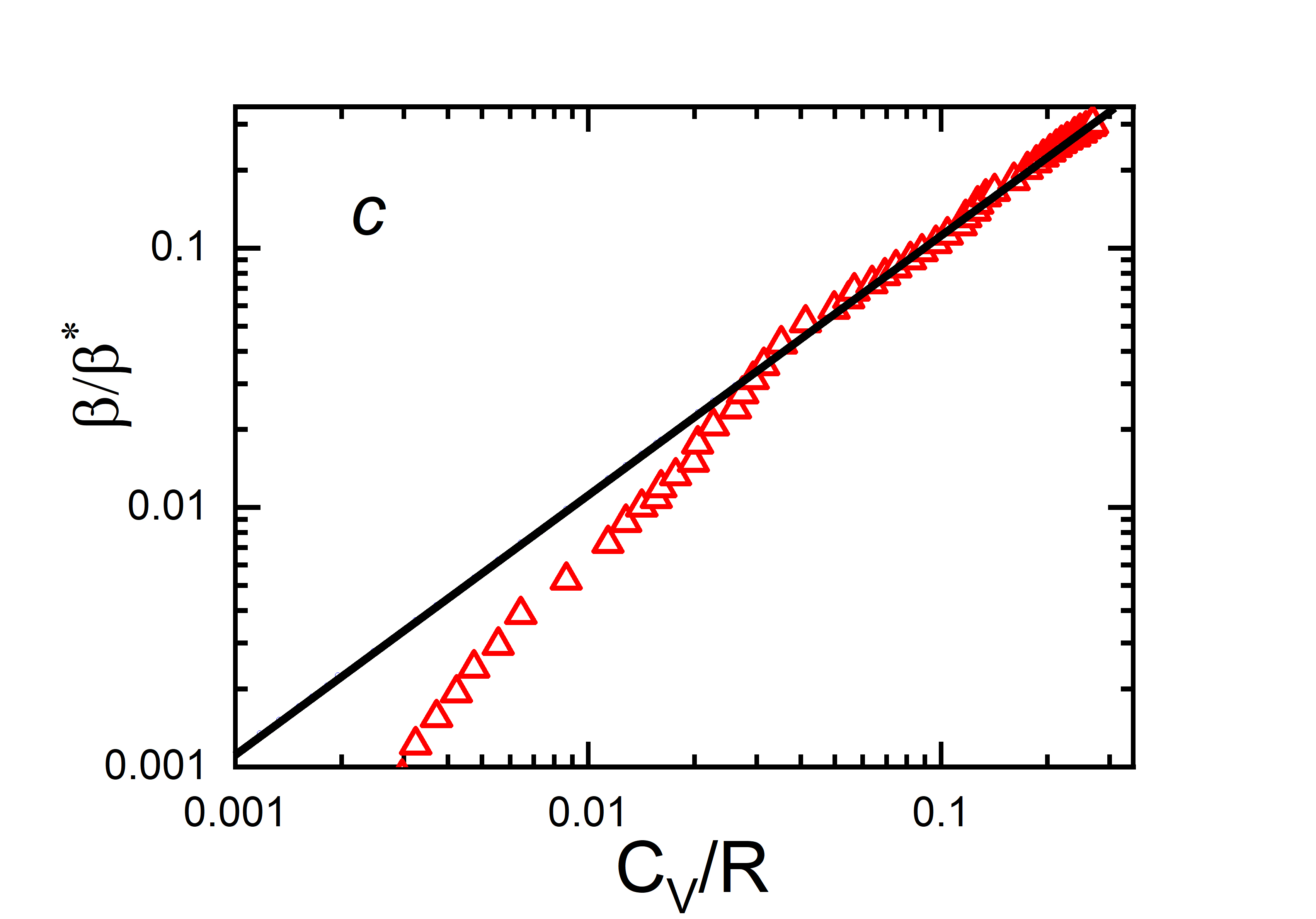} 
	\caption{(Colour online) (a) Temperature dependence $\beta /  \beta^*$  vs. $T$ for compact nanotube system of SWCNTs bundles calculated by using experimental data~\cite{bagatskii2012specific} (red circles); straight line is an approximation by constant  $\beta^* = 9.5\times10^{-5}$~K$^{-1}$;  (b) $C_\textrm{V}(T)/R$ vs. $T$ for compact nanotube system of SWCNTs bundles~\cite{bagatskii2012specific}; (c) Normalized volume thermal expansion $\beta /  \beta^*$ vs. $C_\textrm{V}(T)/R$ for the experimental data~\cite{bagatskii2012specific, dolbin2008vb} of SWCNTs bundles for $\beta^* = 9.5\times10^{-5}$~K$^{-1}$(triangles); the line is a proportional dependence of $\beta /  \beta^*$ vs. $C_\textrm{V}(T)/R$.}
	\label{fig:4}
\end{figure}

Heat capacity at constant pressure $C_\textrm{P}(T)$ of systems of compact nanotubes~\cite{bagatskii2012specific}, which are mecha\-ni\-cal\-ly compressed powder of bundles of single-walled carbon nanotubes (SWCNT), was  studied in the temperature range 2$\div$120 K. The sample for thermal expansion and heat capacity measurements was prepared by compaction under 1 GPa pressure according to Bendiab et al.~\cite{reich2002elastic} method. The molar mass of carbon (12 g$/$mol) was taken to estimate $C/R$ for the system of compacted SWCNT nano\-tubes. The dependence of the volumetric thermal expansion coefficient $\beta \approx 3\alpha$  was estimated using experimental data of linear thermal expansion of the compacted powder of SWCNTs~\cite{dolbin2008vb}. The temperature dependence of the Gr\"uneisen coefficient was calculated by Bagatskii et al.~\cite{bagatskii2012specific} using the value $\chi_\textrm{T} = 2.7\times10^{-11}$~Pa$^{-1}$~\cite{reich2002elastic}, and molar volume $V =1\times10^{-5}$~m$^3/$mol~\cite{bagatskii2012specific}. Temperature dependence $\beta /  \beta^*$ vs. $T$  for the SWCNTs bundles is shown in  figure~\ref{fig:4}a (red circles). It can be seen that this dependence   above 30 K can be approximated by the constant $\beta^* = 9.5\times10^{-5}$~K$^{-1}$  (straight line in figure~\ref{fig:4}a). Note that the system of compacted nanotubes has an extremely low heat capacity, see figure~\ref{fig:4}b). As in the case of $C_\textrm{60}$ fullerite, let us normalize the thermal expansion of the system of compacted nanotubes to the constant $\beta^* = 9.5\times10^{-5}$~K$^{-1}$ (see figure~\ref{fig:4}c, red triangular, experimental data). It is seen that the proportional ratio at $T \geqslant $ 36~K between the volume thermal expansion coefficient and the heat capacity (line in figure~\ref{fig:4}c), that is, a universal linear function $y=kx$ with an angular coefficient $k={\tan} 45^{\circ} = 1$, as well as in the case of the adsorbate system $\beta_\textrm{Xe} /  \beta^*$ vs. $C_\textrm{V}/R$, and $C_\textrm{60}$ fullerite. Deviation of the normalized experimental data from the line is observed at $T < 36$ K, which is explained by a significant decrease of the Gr\"uneisen coefficient~\cite{bagatskii2012specific}. 

Therefore, in both cases of the system of compacted nanotubes SWCNTs and $C_\textrm{60}$, it is possible to propose a generalized linear function $y = kx$, $k=\tan 45^{\circ} = 1$ between $\beta /  \beta^*$ ratio and $C_\textrm{V}/R$, where $\beta^*$ is the normalization parameter for the coefficient of volumetric thermal expansion and R is the universal gas constant. Deviations from this function indicate anomalies in the nature of $\beta /  \beta^*$ on $C_\textrm{V}/R$, which appear for experimental data~\cite{barabashko2017low, bagatskii2015low, gugenberger1992glass}. Previously, the correlation between $\beta$ and $C_\textrm{P}(T)$ for C$_\textrm{60}$ polycrystals was observed in~\cite{bagatskii2015low} at temperatures below 63 K. Above 60 K, the contribution of intramolecular vibrations to the $C_\textrm{P}(T)$ significantly increases, which does not allow a direct comparison between $\beta$ and $C_\textrm{P}(T)$, in particular, to investigate the manifestation of an anomaly characteristic of orientation glass at $T_\textrm{g}$. At the glass transition temperatures $T_\textrm{g}$ (80$\div$90 K) an anomaly is observed in $\alpha(T)$~\cite{gugenberger1992glass}. The proportional correlation $\beta /  \beta^*$ between $C_\textrm{V}/R$ takes place in a wide interval of heat capacity changes up to $C_\textrm{P}/R \approx 4$, which is associated not only with vibrational degrees of freedom $C_\textrm{P}(T)/R \leqslant 3$, but also with the influence of rotational excitations. Next, the correlation between heat capacity and thermal expansion in atomic and simple molecular crystals will be considered to reveal the influence of rotational degrees of freedom.

\subsection{Ratio between volume thermal expansion $\beta$ and the heat capacity $C_\textrm{V}$ of atomic and simple molecular crystals}

\begin{figure}[!htb]
	\centering
	\includegraphics[width=0.45\linewidth]{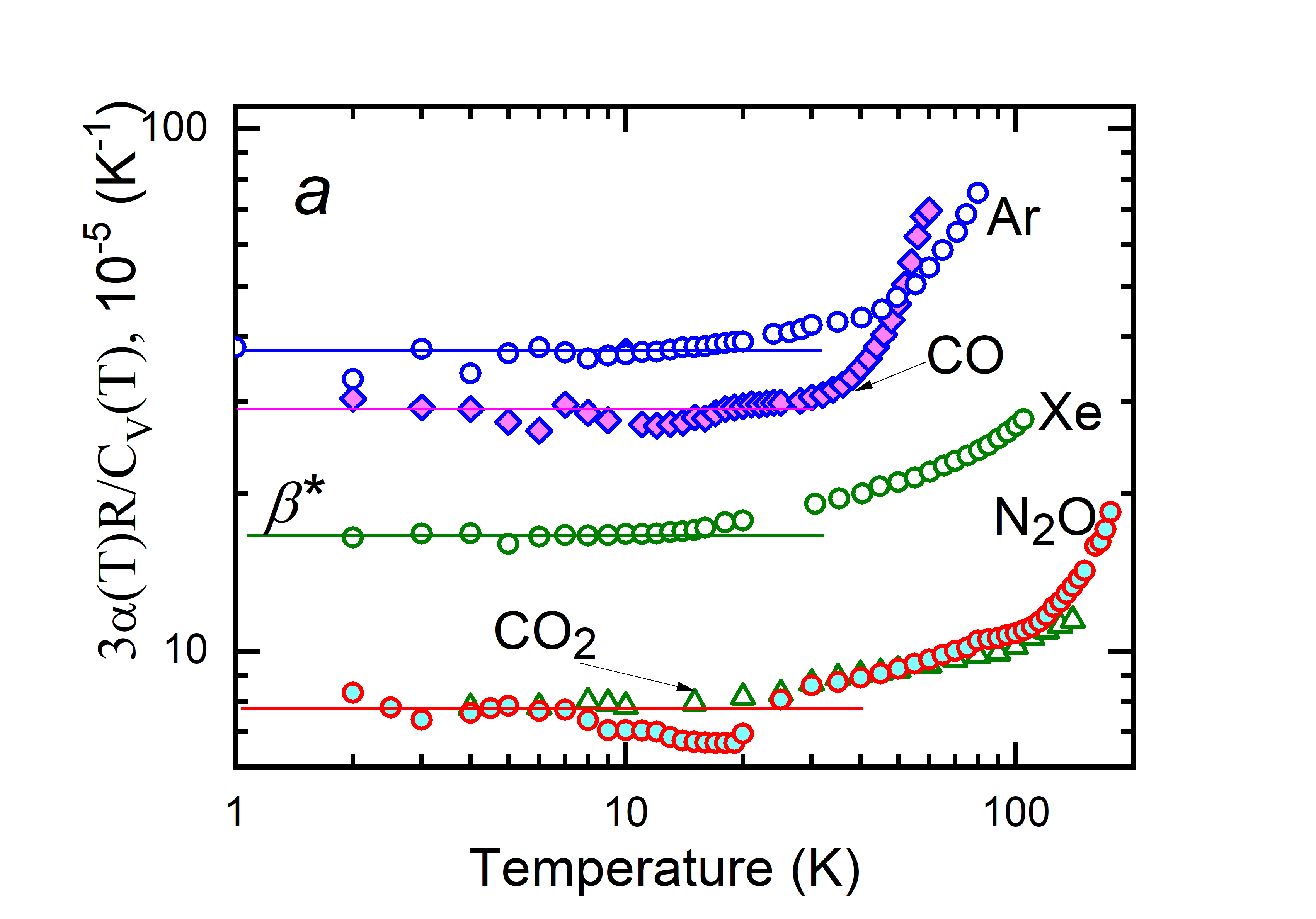}  
	\includegraphics[width=0.45\linewidth]{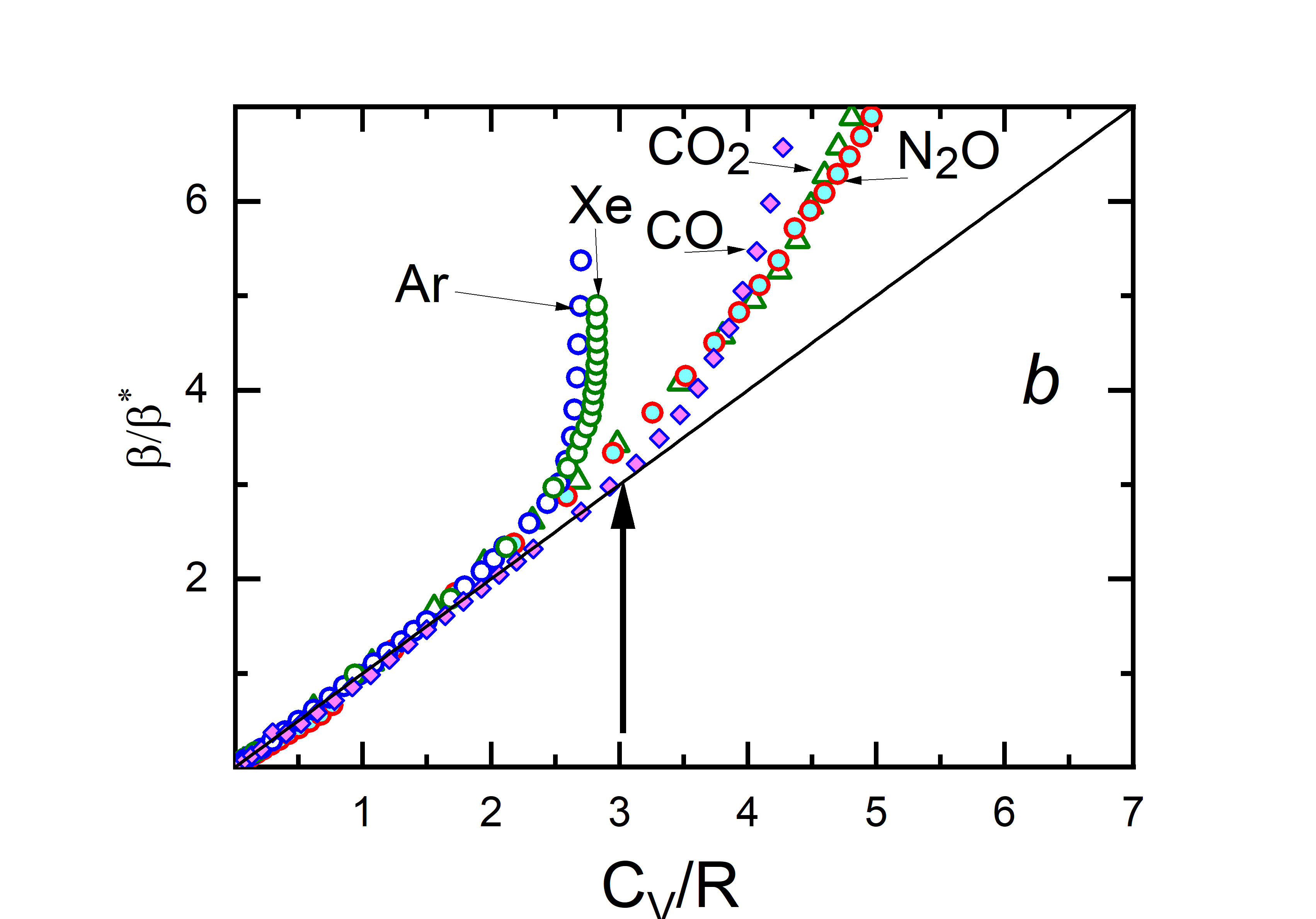}  
	\caption{(Colour online) (a) The calculated temperature dependence $\beta^*$ according to the equation 2.4  vs. T for Ar, Xe,  CO$_{2}$, CO, N$_{2}$O (symbols); solid straight lines are  an approximation, constant $\beta^*$, see table~\ref{tab:1}. (b) Normalized volume thermal expansion $\beta/\beta^*$ vs. $C_\textrm{V}/R$ (symbols are the same as in figure~\ref{fig:5}a): N$_\textrm{2}$O (data~\cite {gavrilko1999structure, manzheliui1997physics, venables1977rare}   and $\beta^{*} = 8\times10^{-5}$~K$^{-1}$), CO$_\textrm{2}$ (data~\cite {gavrilko1999structure, manzheliui1997physics, venables1977rare} and $\beta^* = 8\times10^{-5}$~K$^{-1}$), CO (data~\cite {gavrilko1999structure, manzheliui1997physics, venables1977rare} and $\beta^* = 30\times10^{-5}$~K$^{-1}$), solid Xe (data~\cite {gavrilko1999structure, manzheliui1997physics, venables1977rare} and $\beta^* = 16\times10^{-5}$~K$^{-1}$); solid Ar (data~\cite {gavrilko1999structure, manzheliui1997physics, venables1977rare} and $\beta^* = 38\times10^{-5}$~K$^{-1}$). The line is a direct proportional dependence of $\beta/\beta^*$ on $C_\textrm{V}/R$.}
	\label{fig:5}
\end{figure}

Figure~\ref{fig:5}a shows the calculated temperature dependence   $\beta^*$ vs. $T$ for cryocrystals Ar, Xe, CO$_{2}$, CO, N$_{2}$O (symbols). It can be seen that both for atomic cryocrystals and for cryocrystals with linear symmetry the value   $\beta^*$  vs. $T$ is close to the corresponding  $\beta^*$  (solid straight lines). The values of  $\beta^*$  are given in table~\ref{tab:1}.
Figure~\ref{fig:5}b shows that the correlation described by a universal linear function with the angular coefficient $k = \tan 45^{\circ} = 1$ between the normalized thermal expansion ($\beta /  \beta^*$) and the heat capacity normalized to $R$ ($C_\textrm{V}/R$), was observed for the experimental data of atomic crystals, such as Xe, Ar, that have 3 degrees of freedom, and the maximum heat capacity of the crystal is equal to $3R$. For atomic crystals, such a linear correlation was observed in the entire range of temperatures from the lowest experimental to temperatures where $C_\textrm{V}/R \approx 2.3$. In the temperature range, where $2.3 < C_\textrm{V}/R < 3$ (the classical limit of Dulong and Petit) a sharp increase of the function ($\beta /\beta^{*}$) is observed which means the disappearance of the correlation between volumetric thermal expansion and heat capacity. Note that atomic crystals (Xe, Ar)  have three independent acoustic branches in the dispersion law. For molecular crystals with linear molecules, such as CO$_{2}$, CO, N$_{2}$O, the deviation from the linear dependence $\beta /  \beta^*$ vs. $C_\textrm{V}/R$  is observed at values of heat capacity $C_\textrm{V}/R \approx 3\div3.5$. For molecular crystals, in addition to the translational degrees of freedom, there are degrees of freedom associated with the rotational motion of the molecules (2 additional degrees of freedom are added in the case of CO$_{2}$, CO, N$_{2}$O). This means that linear correlation ($\beta /  \beta^*$) vs. $C_\textrm{V}/R$  is associated not only with translational degrees of freedom but also with rotational degrees of freedom. It is seen that for more degrees of freedom, the linear correlation is observed in the wider interval. This fact indicates that such correlation for crystals depends on the number of degrees of freedom of the molecules, and not on the nature of the atoms or molecules themselves. 

\begin{figure}[!htb]
	\centering
	\includegraphics[width=0.45\linewidth]{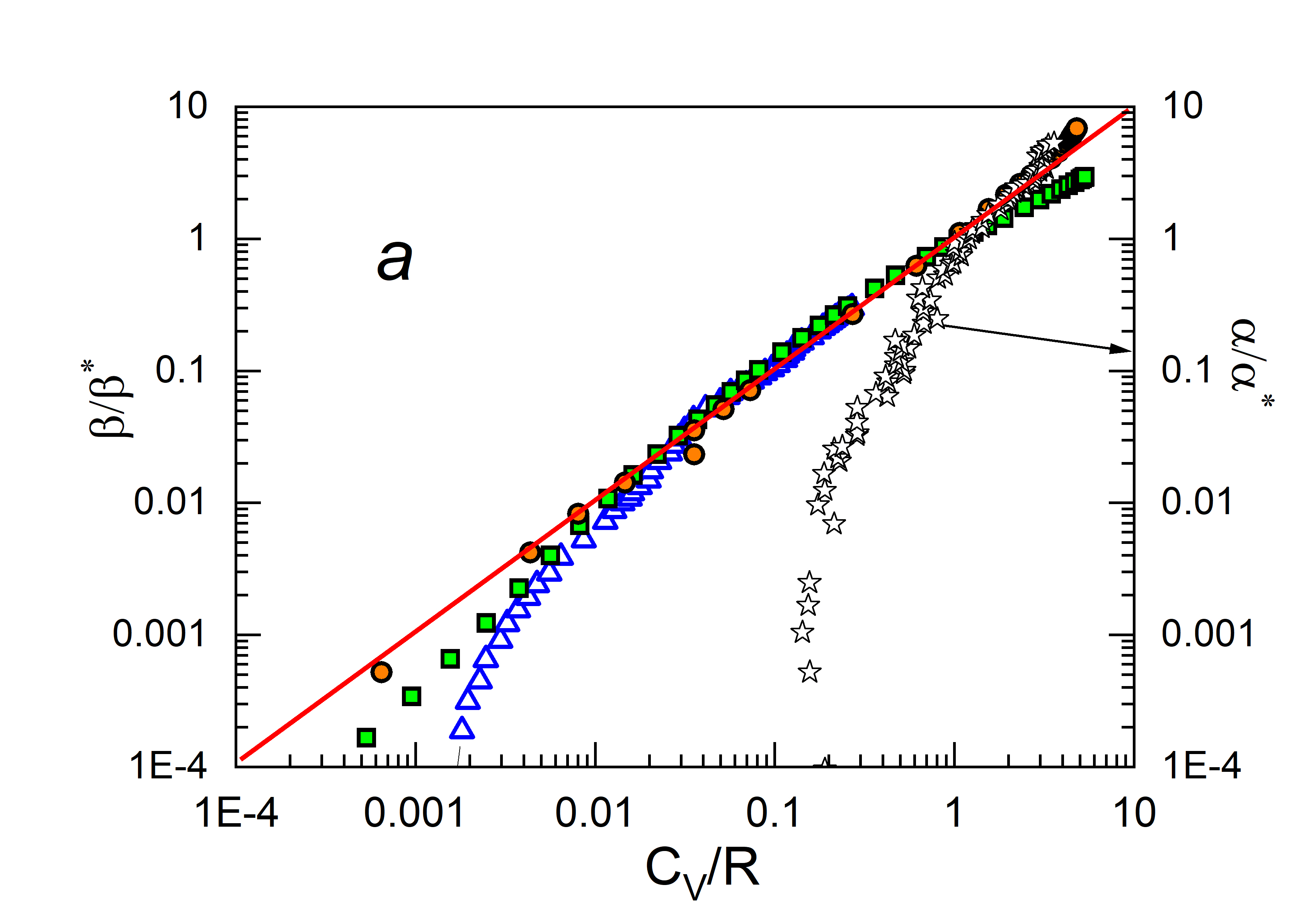}  
	\includegraphics[width=0.45\linewidth]{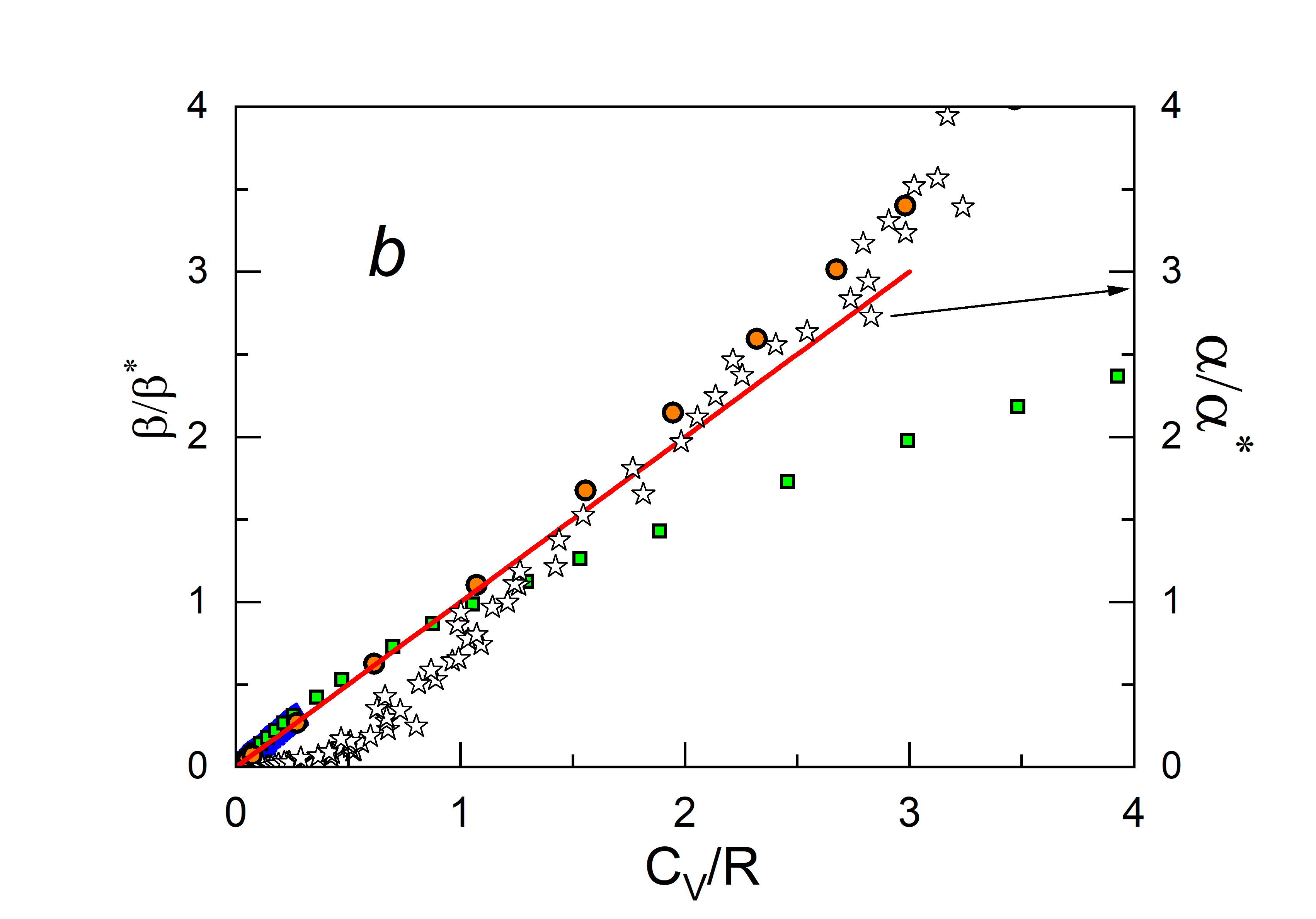}  
	\caption{(Colour online) Normalized volume thermal expansion  $\beta/\beta^*$ vs. $C_\textrm{V}/R$ (6a --- in double logarithmic coordinates, 6b --- in linear coordinates): contribution of Xe atoms adsorbed by the system of compacted carbon nanotubes ($\alpha^* = 0.1\times10^{-5}$~K$^{-1}$) (stars); CO$_\textrm{2}$ ($\bullet$, data~\cite {gavrilko1999structure, manzheliui1997physics, venables1977rare} and $\beta^*=8\times10^{-5}$~K$^{-1}$); SWCNTs bundles ($\triangle$) plotted for experimental data~\cite {bagatskii2012specific, dolbin2008vb}  and $\beta^* = 9.5\times10^{-5}$~K$^{-1}$; amorphous SiO$_2$ ($\square$, data~\cite {barron1982thermal}  and $\beta^* = 1.2\times10^{-5}$~K$^{-1}$). }
	\label{fig:6}
\end{figure}

In figure~\ref{fig:6}, the values of $\beta/\beta^*$ from $C_\textrm{V}/R$ for amorphous SiO$_2$ are added for comparison, since amorphous SiO$_{2}$ belongs to the class of amorphous materials~\cite {barron1982thermal}. Furthermore, for comparison, figure~\ref{fig:6} shows the dependence $\alpha_{\rm Xe}/\alpha^*$ vs. $C_\textrm{P}/R$ of the contribution of Xe adsorbed by the system of compacted carbon nanotubes is normalized to $\alpha^*$ and the dependence of $\beta/\beta^*$ on the $C_\textrm{V}/R$ of the system of mechanically compacted SWCNTs and CO$_{2}$. In figure~\ref{fig:6} the dependence of $\beta/\beta^*$ vs. $C_\textrm{V}/R$ is presented in a double logarithmic coordinate system that better represents the range of low temperatures. For CO$_{2}$, as well as for other atomic and molecular crystals, up to the lowest temperatures, a linear dependence of $\beta/\beta^*$on $C_\textrm{V}/R$ is observed. As for amorphous SiO$_{2}$, as in the case of SWCNTs and the contribution of Xe atoms to the SWCNT--Xe system, there is a deviation from a linear dependence with decreasing temperature, which characterizes the presence of a negative contribution to thermal expansion, that leads to a decrease and even negative values of the Gr\"uneisen parameter $\beta(T)$ of materials with the structural disorder. Figure~\ref{fig:6}b shows the high-temperature range of $\beta/\beta^*$ from $C_\textrm{V}/R$. In the case of amorphous SiO$_{2}$, the deviation from the linear dependence $\beta/\beta^*$ from $C_\textrm{V}/R$ occurs at $C_\textrm{V}/R \geqslant 1$, which indicates that the  volume thermal expansion and heat capacity become independent of each other.

\begin{table}[!htb]
\caption{\label{tab:1} The normalization parameter for the coefficient of linear thermal expansion is $\alpha^*$ for the case of Xe atoms adsorbed by a system of compacted carbon nanotubes, and the normalization parameter for the coefficient of volumetric thermal expansion is $\beta^*$ for the case of atomic crystals, molecular crystals and disordered systems.}
\begin{center}
\begin{tabular}{||c c c c ||} 
 \hline
 Material & $\alpha^*$; $\beta^*$ & References for   & References for the  \\
 & & $C(T)$ data & expansion data \\

 \hline
 
  Atomic crystals &  &  &\\
 solid Ar  & $\beta^* = 38\times10^{-5}$ K$^{-1}$  & \cite {gavrilko1999structure, manzheliui1997physics, venables1977rare} & \cite {gavrilko1999structure, manzheliui1997physics, venables1977rare}  \\
  solid Xe  & $\beta^* = 16\times10^{-5}$ K$^{-1}$  & \cite {gavrilko1999structure, manzheliui1997physics, venables1977rare} & \cite {gavrilko1999structure, manzheliui1997physics, venables1977rare}  \\
   \hline
  Molecular crystals &  &  &\\
 solid CO$_2$  & $\beta^* = 8\times10^{-5}$ K$^{-1}$  & \cite {gavrilko1999structure, manzheliui1997physics, venables1977rare} & \cite {gavrilko1999structure, manzheliui1997physics, venables1977rare}  \\
 solid CO  & $\beta^* = 30\times10^{-5}$ K$^{-1}$  & \cite {gavrilko1999structure, manzheliui1997physics, venables1977rare} & \cite {gavrilko1999structure, manzheliui1997physics, venables1977rare}  \\
 solid N$_2$O  & $\beta^* = 8\times10^{-5}$ K$^{-1}$  & \cite {gavrilko1999structure, manzheliui1997physics, venables1977rare} & \cite {gavrilko1999structure, manzheliui1997physics, venables1977rare}  \\
  \hline
  Strong anisotropic solids &  &  &\\
 \hline
 adsorbed Xe  & $\alpha^* = 0.1\times10^{-5}$ K$^{-1}$  & \cite {bagatskii2013experimental} & \cite{dolbin2009radial}  \\
  \hline
  SWCNTs bundles  & $\beta^* = 9.5\times10^{-5}$ K$^{-1}$  & \cite {bagatskii2012specific} & \cite{dolbin2008vb}  \\
  \hline
   Disordered solids &  &  &\\
    \hline
   solid C$_{60}$  & $\beta^* = 0.65\times10^{-5}$ K$^{-1}$  & \cite {barabashko2017low, bagatskii2015low} & \cite{aleksandrovskii2005polyamorphism, gugenberger1992glass}  \\
\hline
   solid SiO$_2$
  & $\beta^* = 1.2\times10^{-5}$ K$^{-1}$  & \cite {barron1982thermal} & \cite {barron1982thermal}  \\
 \hline
\end{tabular}

\end{center}
\end{table}

It can be seen that in the same way as in the case of SWCNTs and the system of compacted carbon nanotubes with adsorbed Xe atoms, the structural disorder of amorphous SiO$_\textrm{2}$ leads to the negative contribution to thermal expansion~\cite {barron1982thermal, miller2009negative}, and the linear dependence $(\beta/\beta^*)$ on $(C_\textrm{V}/R)$  is violated (see figure~\ref{fig:6}) at lower temperatures. For carbon nanomaterials, as well as for atomic and simple molecular crystals, a linear ratio between volumetric thermal expansion and heat capacity is observed only in the temperature range at which the relative change in atomic vibration frequencies to the change of volume is constant.

To explain the proportional correlation observed for simple atomic and molecular crystals, we analyzed new experimental results of the universal behaviour of the low-temperature heat capacity of molecular crystals~\cite{krivchikov2022role, szewczyk2021heat,miyazaki2021low}. It was found that all frequencies of the real spectrum are proportional to the frequency $\omega_\textrm{vH}$ of the first van Hove singularity, the magnitude of which depends on the volume. Therefore, the Gr\"uneisen parameter becomes the same for all modes~\cite{krivchikov2022role, szewczyk2021heat,miyazaki2021low}.

The universal behaviour of the temperature dependence of the experimental heat capacity  $C_\textrm{exp}(T)/T^3$, which has a maximum at the temperature $T_\textrm{max}$ follows as a result of the first van Hove feature in the density of vibrational states. The frequency $\omega_\textrm{vH} \approx 5T_\textrm{max}k_{\textrm{B}}/\hbar$, where $k_{\textrm{B}}$ and $\hbar$ are the Boltzmann and Planck constants, respectively~\cite{strzhemechny2019heat}.

The normalized function of excess $\Delta C = [C_\textrm{exp}(T) - C_\textrm{D}(T)]/T^3$ of the experimental values of heat capacity $C_\textrm{exp}(T)$ over the Debye contribution $C_\textrm{D}(T) = c_\textrm{3}T^3$ ($c_\textrm{3}$ --- coefficient) vs. $T/T_\textrm{max}$, in a wide temperature range has a universal dependence for ordered and disordered materials~\cite{krivchikov2022role}:
\begin{equation}\label{8}
 \Delta^*C = \frac{\left[{C_\textrm{exp}(T)}/{T^3}\right] - c_\textrm{3}}{\left[{C_\textrm{exp}(T)}/{T^3}\right ]_\textrm{max} - c_\textrm{3}}.
\end{equation}
Therefore, the Gr\"uneisen parameter:
\begin{equation}\label{9} 
\gamma = -\frac{\partial (\ln\omega)}{\partial (\ln V)} = -\frac{\partial (\ln\omega_\textrm{vH})}{\partial (\ln V)} \approx -\frac{\partial (\ln\theta_\textrm{vH})}{\partial (\ln V)},
\end{equation}
where $\Theta_\textrm{vH} \approx 5T_\textrm{max}$~\cite{strzhemechny2019heat}.

\begin{figure}[!htb]
	\centering
	\includegraphics[width=0.9\linewidth]{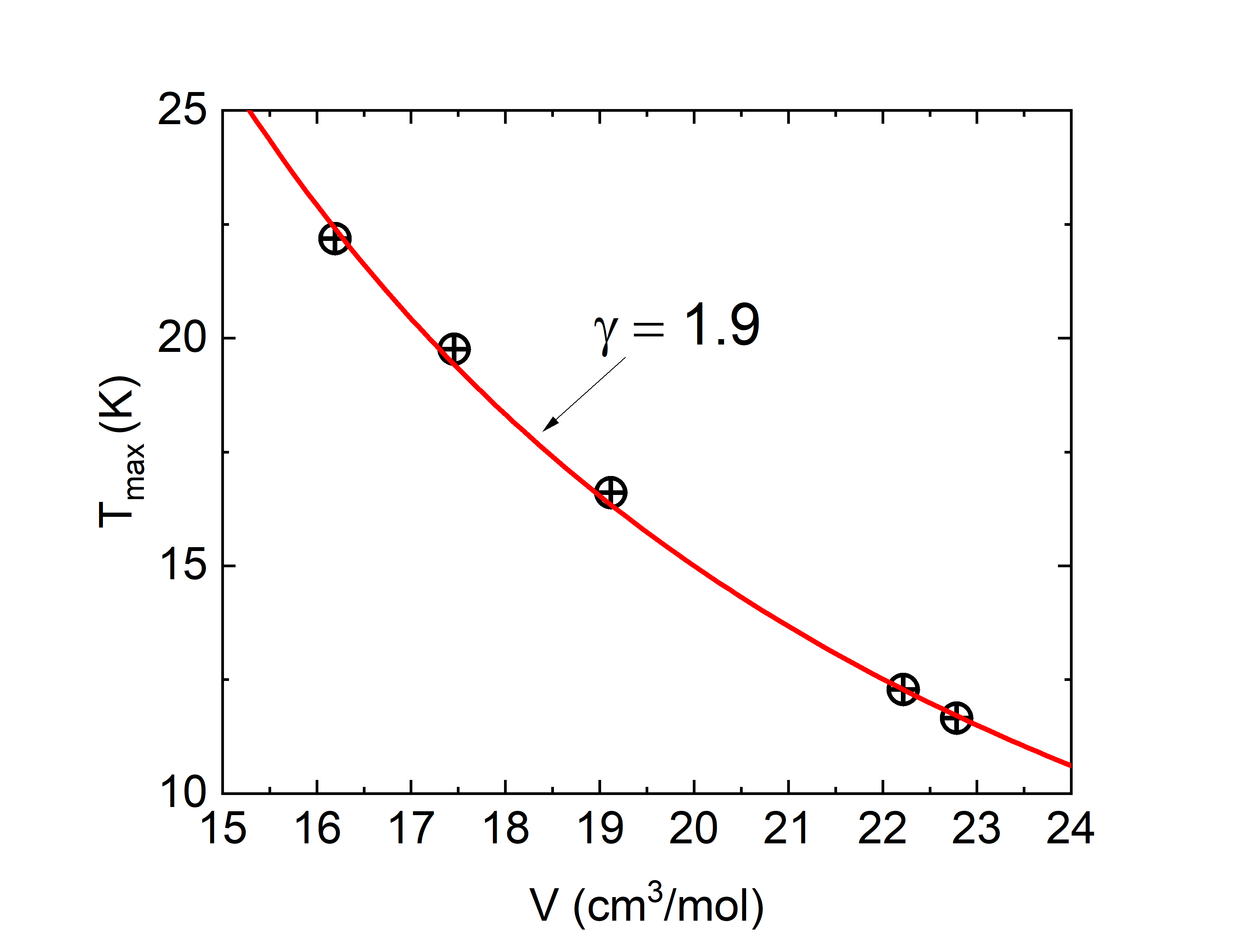}  
		\caption{ (Colour online) Dependence of the temperature $T_\textrm{max}$ of the maximum heat capacity $C_\textrm{exp}(T)/T^3$ vs. the molar volume according to experimental data~\cite{krause1980direct}: symbols are experimental data; solid curve is equation~(\ref{10}) for $\gamma = 1.9$.}
	\label{fig:7}
\end{figure}

To verify the equations~(\ref{8} and \ref{9}), we used experimental data of the heat capacity for solid parahydrogen at constant volume in the range of molar volumes from 22.79 to 16.19~cm$^3$/mol~\cite{krause1980direct}. The dependence of the temperature $T_{\rm max}$ vs. the molar volume is given in table~\ref{tab:2} and is shown in figure~\ref{fig:7}. 

\begin{table}[!htb]
\caption{\label{tab:2} Dependence of the temperature $T_{\rm max}$ of the maximum heat capacity $C_\textrm{exp}(T)/T^3$ vs. the molar volume according to experimental data~\cite{krause1980direct}. }
\begin{center}
\begin{tabular}{||c c ||} 
 \hline 
 $V$ & $T_{\rm max}$, in $C_\textrm{exp}(T)/T^3$ \\[0.3ex] 
 cm$^3$/mol  & K \\[0.3ex]
 \hline
 22.221 &	12.28 \\  \hline
22.787 &	11.65 \\   \hline
19.120 &	16.6 \\  \hline
17.458 &	19.75 \\    \hline
16.193 &	22.18 \\   
 \hline
\end{tabular}

\end{center}
\end{table}

Several conclusions emerge from the analysis of the data in figure~\ref{fig:7} and table~\ref{tab:2}. First, it was found that the dependence $\Delta^*C(T/T_\textrm{max})$ is universal in a wide range of normalized temperatures $T/T_\textrm{max}$ from 0 to 2.5. Second, it was obtained that the value of the Gr\"uneisen parameter is equal to $\gamma = 1.9$, and the experimental data agree well with the dependence:
\begin{equation}\label{10} 
 \frac{\Theta_\textrm{vH}(V_\textrm{1})}{\Theta_\textrm{vH}(V_\textrm{2})} = \frac{T_\textrm{max}(V_\textrm{1})}{T_\textrm{max}(V_\textrm{2})} = \left(\frac{V_\textrm{2}}{V_\textrm{1}}\right )^\gamma.
\end{equation}
Note that the facts discussed above  can explain the proportional correlation between heat capacity and thermal expansion for crystals with appear the first Van Hove feature in the density of vibrational states. In the case of strongly anisotropic solids, the first van Hove feature is not realized in the density of vibrational states, and in this case there is no condition for realizing a proportional correlation at lower temperatures between heat capacity and thermal expansion. 

\section{Conclusions}

The relationship between the heat capacity $C(T)$ and the thermal expansion for Xe atoms adsorbed by a system of compacted carbon nanotubes, which are mechanically compressed bundles of single-walled carbon nanotubes, single crystals, and amorphous materials, was analyzed. The proportional correlation between the normalized volume thermal expansion $(\beta/\beta^*)$ and the heat capacity $(C_\textrm{V}/R)$, is observed for atomic cryocrystals such as Xe, Ar, in the temperature range from the lowest experimental to temperatures where $(C_\textrm{V}/R) \approx 2.3$. This correlation depends on the number of vibration modes and does not depend on the characteristics of the atoms of these crystals. In the temperature region, where $2.3 < C_\textrm{V}/R < 3$ (the classical limit of Dulong and Petit) the correlation  disappears, which indicates that the volumetric thermal expansion and heat capacity become independent of each other. For molecular crystals with linear symmetry, such as CO$_{2}$, CO, N$_{2}$O, a universal proportional ratio between heat capacity and thermal expansion is also observed, and the deviation from the linear dependence $(\beta/\beta^*)$ on $(C_\textrm{V}/R)$  is observed at values of heat capacity $(C_\textrm{V}/R) \approx 3 \div 3.5$. These facts indicate that the proportional correlation is related not only to the translational but also to the rotational degrees of freedom of the molecule in the crystal. In the case of the C$_{60}$ molecular crystal with translational, rotational degrees of freedom and intermolecular vibrations, the above-discussed correlation occurs up to the values of heat capacity $(C_\textrm{V}/R) \approx 7.5$. In strongly anisotropic systems, such as systems of compacted carbon nanotubes, which are mechanically compressed bundles of SWCNTs and systems of Xe atoms adsorbed by compacted carbon nanotubes, this universal dependence is violated at low temperatures. A qualitative explanation of the observed correlation is proposed.

\section*{Acknowledgements}

The study was supported by the National Research Foundation of Ukraine (Grant No.197/02.2020).



\begin{thebibliography}{99}
		
\bibitem{white1993solids} White G.~K., Contemp. Phys., 1993, \textbf{34}, No.~4, 193--204, \doi{10.1080/00107519308213818}.
\bibitem{gavrilko1999structure} Gavrilko~V.~G., Isakina A., Manzhelii~V.~G., Prokhvatilov A., Structure and Thermodynamic Properties of Cryocrystals Handbook, Begell House, 1999.
\bibitem{hassaine2012low} Hassaine M., Ramos M. A., Krivchikov A.~I., Sharapova I.~V., Korolyuk O.~A., Jim\'enez-Riob\'oo R.~J., Phys.~Rev.~B, 2012, \textbf{85}, No.~10, 104206, \doi{10.1103/PhysRevB.85.104206}.
\bibitem{manzhelii1998thermodynamic} Manzhelii V.~G., Bagatskii M.~I., Minchina I.~Ya., Aleksandrovskii A.~N., J. Low Temp. Phys., 1998, \textbf{111}, No.~3, 257--270, \doi{10.1023/A:1022219232478}.
\bibitem{barabashko2021calorimetric} Barabashko M.~S., Drozd M., Szewczyk D., Je\.zowski A., Bagatskii~M.~I., Sumarokov V.~V., Dolbin~A.~V., Nesov~S.~N., Korusenko~P.~M., Ponomarev~A.~N., Geidarov~V.~G., Kuznetsov~V.~L., Moseenkov~S.~I., Sokolov~D.~V., Smirnovi~D.~A., Fullerenes,~Nanotubes~Carbon~Nanostruct., 2021, \textbf{29}, No.~5, 331--336, \doi{10.1080/1536383X.2020.1819251}.
\bibitem{bagatskii2021size} Bagatskii M.~I., Je\.zowski A., Szewczyk D., Sumarokov V.~V., Barabashko~M.~S., Kuznetsov~V.~L., Moseenkov~S.~I., Ponomarev~A.~N., Therm.~Sci.~Eng.~Prog., 2021, \textbf{26}, 101097, \doi{10.1016/j.tsep.2021.101097}.
\bibitem{barabashko2017low} Barabashko M.~S., Rezvanova A.~E., Ponomarev A.~N., Fullerenes,~Nanotubes~Carbon~Nanostruct., 2017, \textbf{25}, No.~11, 661--666, \doi{10.1080/1536383X.2017.1391225}.
\bibitem{dolbin2019thermal} Dolbin A.~V., Khlistuck M.~V., Eselson V.~B., Gavrilko V.~G., Vinnikov N.~A., Basnukaeva R.~M., Konstantinov~V.~A., Luchinskii~K.~R., Nakazawa~Y., Low~Temp.~Phys., 2019, \textbf{45}, No.~1, 128--131, \doi{10.1063/1.5082324}.
\bibitem{bagatskii2014low} Bagatskii M.~I., Manzhelii V.~G., Sumarokov V.~V., Dolbin~A.~V., Barabashko~M.~S., Sundqvist~B., Low~Temp.~Phys., 2014, \textbf{40}, No.~8, 678--684, \doi{10.1063/1.4892643}.
\bibitem{vinnikov2022analysis} Vinnikov N.~A., Cherednichenko~S.~V., Dolbin~A.~V., Eselson~V.~B., Gavrilko~V.~G., Basnukaeva~R.~M., Plokhotnichenko~A.~M., Low Temp. Phys., 2022, \textbf{48}, No.~4, 336--338, \doi{10.1063/10.0009739}.
\bibitem{dolbin2020influence} Dolbin~A.~V., Dubinko~V.~I., Vinnikov N.~A., Esel’son~V.~B., Gavrilko~V.~G., Basnukaeva~R.~M., Khlistyuck~M.~V., Cherednichenko~S.~V., Kotsyubynsky~V.~O., Boychuk~V.~M., Kolkovsky~P.~I., Low~Temp.~Phys., 2020, \textbf{46}, No.~10, 1030--1038, \doi{10.1063/10.0001921}.
\bibitem{krivchikov2022role} Krivchikov A. I., Je\.zowski A., Szewczyk D., Korolyuk O. A., Romantsova O. O., Buravtseva L. M., Cazorla C., Tamarit J. L., J. Phys. Chem. Lett., 2022, \textbf{13}, 5061--5067, \doi{10.1021/acs.jpclett.2c01224}.
\bibitem{rusakova2020possible} Rusakova H.~V., Fomenko L.~S., Lubenets S.~V., Dolbin~A.~V., Vinnikov~N.~A., Basnukaeva~R.~M., Khlistyuck~M.~V., Blyznyuk~A.~V., Low Temp. Phys., 2020, \textbf{46}, No.~3, 276--284, \doi{10.1063/10.0000699}.
\bibitem{manzhelii2017influence} Manzhelii~E.~V., J. Low Temp. Phys., 2017, \textbf{187}, No.~1, 105--112, \doi{10.1007/s10909-016-1699-1}.
\bibitem{bagatskii2017heat} Bagatskii~M.~I., Barabashko~M.~S., Sumarokov~V.~V., Je\.zowski~A., Stachowiak~P., J. Low Temp. Phys., 2017, \textbf{187}, No.~1, 113--123, \doi{10.1007/s10909-016-1737-z}.
\bibitem{dolbin2017thermal} Dolbin~A.~V., Khlistyuck~M.~V., Eselson~V.~B., Gavrilko~V.~G., Vinnikov N.~A., Basnukaeva~R.~M., Concei\c{c}\~{a}o~F., Ochoa~M., J. Appl. Phys. Sci. Int., 2017, \textbf{8}, 47--52.
\bibitem{bagatskii2016heat} Bagatskii~M.~I., Sumarokov~V.~V., Barabashko~M.~S., Low Temp. Phys., 2016, \textbf{42}, No. 2, 94--98,\\ \doi{10.1063/1.4942395}.
\bibitem{manzhelii1971thermal} Manzhelii~V.~G., Tolkachev A.~M., Bagatskii~M.~I., Voitovich~E.~I., Phys. Status Solidi B, 1971, \textbf{44}, No.~1, 39--49,\\
\doi{10.1002/pssb.2220440104}.
\bibitem{barron2012heat} Barron T. H. K., White G. K., Heat Capacity and Thermal Expansion at Low Temperatures, Springer~Science~\&~Business~Media, New York, 2012.
\bibitem{ramos2013low} Ramos M. A., Hassaine M., Kabtoul B., Jim\'enez-Riob\'oo R., Shmyt’ko I.~M., Krivchikov~A.~I., Sharapova~I.~V., Korolyuk~O.~A., Low Temp. Phys., 2013, \textbf{39}, No. 5, 468--472, \doi{10.1063/1.4807147}.
\bibitem{gruneisen1912theorie} Gr\"uneisen E., Ann. Phys., 1912, \textbf{344}, No. 12, 257--306, \doi{10.1002/andp.19123441202}.
\bibitem{pathak2022thermal} Pathak K. N., Deo B., Phys. Status Solidi B, 1966, \textbf{17}, No.~1, 77--82,
\doi{10.1002/pssb.19660170110}.
\bibitem{meincke1962two} Meincke P. P. M., Can. J. Phys., 1962, \textbf{40}, No.~2, 283--285, \doi{10.1139/p62-025}.
\bibitem{leibfried1961theory} 
Leibfried~G., Ludwig~W., In: Solid {{State Physics}}, Vol.~12, Seitz~F.,
Turnbull~D. (Eds.), {Academic Press}, 1961, 275--444,
\doi{10.1016/S0081-1947(08)60656-6}.
\bibitem{tang2021scaling} Tang M., Pan X., Zhang M., Wen H., Chin. Phys. Lett., 2021, \textbf{38}, No. 2, 026501, \doi{10.1088/0256-307X/38/2/026501}.
\bibitem{bodryakov2014correlation} Bodryakov~V.~Yu., Phys. Solid State, 2014, \textbf{56}, No.~11, 2359--2365, \doi{10.1134/S1063783414110043}.
\bibitem{bodryakov2015correlation} Bodryakov~V.~Yu., High Temp., 2015, \textbf{53}, No.~5, 643--648, \doi{10.1134/S0018151X15040069}.
\bibitem{manzheliui1997physics} Manzhelii~V.~G., Freiman Y. A., Physics of Cryocrystals, American Institute of Physics, New York, 1997.
\bibitem{venables1977rare} Klein M. L., Venables J., Rare Gas Solids, Academic Press, 1977.
\bibitem{bagatskii2012specific} Bagatskii~M.~I., Barabashko~M.~S., Dolbin A. V., Sumarokov~V.~V., Sundqvist B., Low Temp. Phys., 2012, \textbf{38}, No.~6, 523--528, \doi{10.1063/1.4723677}.
\bibitem{schelling2003thermal} Schelling P. K., Keblinski P., Phys. Rev. B, 2003, \textbf{68}, No.~3, 035425, \doi{10.1103/PhysRevB.68.035425}.
\bibitem{bagatskii2015low} Bagatskii~M.~I., Sumarokov~V.~V., Barabashko~M.~S., Dolbin A. V., Sundqvist B., Low Temp. Phys., 2015, \textbf{41}, No.~8, 630--636, \doi{10.1063/1.4928920}.
\bibitem{bagatskii2013experimental} Bagatskii~M.~I., Manzhelii~V.~G., Sumarokov~V.~V., Barabashko~M.~S., Low Temp. Phys., 2013, \textbf{39}, No.~7, 618--621, \doi{10.1063/1.4816120}.
\bibitem{barabashko2015heat} Barabashko~M.~S., Bagatskii~M.~I., Sumarokov~V.~V., In: Nanotechnology in the Security Systems. NATO Science for Peace and Security Series C: Environmental Security, Bon\v{c}a~J., Kruchinin~S. (Eds.),  Springer, Dordrecht, 2015, 121--130, \doi{10.1007/978-94-017-9005-5_11}.
\bibitem{bagatskii2014thermal} Bagatskii~M.~I., Barabashko~M.~S., Sumarokov~V.~V.,	JETP Lett., 2014, \textbf{99}, No.~8, 461--465,\\
\doi{10.1134/S0021364014080049}.
\bibitem{barabashko2021experimental} Barabashko~M.~S., Nakazawa Y., Netsu Sokutei, 2021, \textbf{48}, No.~4, 164--170, \doi{10.11311/jscta.48.4_164}.
\bibitem{dolbin2009radial} Dolbin A. V., Esel'son~V.~B., Gavrilko~V.~G., Manzhelii V. G., Popov S. N, Vinnikov N.~A., Danilenko~N.~I., Sundqvist~B., Low Temp. Phys., 2009, \textbf{35}, No.~6, 484--490, \doi{10.1063/1.3151995}.
\bibitem{dolbin2008vb} Dolbin A. V., Esel'son V. B.,  Gavrilko V. G., Manzhelii~V.~G., Vinnikov~N.~A., Popov~S.~N., Sundqvist~B., Low~Temp.~Phys., 2008, \textbf{34}, 678, \doi{10.1063/1.2967518}.
\bibitem{manzhelii2015phonon} Manzhelii~E.~V., Feodosyev S.~B., Gospodarev I.~A., Syrkin E.~S., Minakova~K.~A., Low Temp. Phys., 2015, \textbf{41}, No.~7, 557--562, \doi{10.1063/1.4927047}.
\bibitem{hone2000quantized} Hone J., Batlogg B., Benes Z., Johnson A.~T., Fischer~J.~E., Science, 2000, \textbf{289}, No.~5485, 1730--1733,\\ \doi{10.1126/science.289.5485.1730}.
\bibitem{landau2013statistical} Landau~L.~D., Lifshits~E.~M., Statistical physics, Course of theoretical physics, Vol.~5, {Elsevier}, 3 edn., 2013.
\bibitem{gruneisen1908zusammenhang} Gr\"uneisen E., Ann. Phys., 1908, \textbf{331}, No.~7, 393--402, \doi{10.1002/andp.19083310707}, (in German).
\bibitem{garai2006correlation} Garai J., Calphad, 2006, \textbf{30}, No.~3, 354--356, \doi{10.1016/j.calphad.2005.12.003}.
\bibitem{aksenova1999analysis} Aksenova N. A., Isakina A. P., Prokhvatilov A. I., Strzhemechny M. A., Low Temp. Phys., 1999, \textbf{25}, No.~8, 724--731, \doi{10.1063/1.593803}.
\bibitem{aleksandrovskii2005polyamorphism} Aleksandrovskii A. N., Dolbin~A.~V., Esel'son~V.~B., Gavrilko~V.~G., Manzhelii~V.~G., Bakai A. S., Cassidy~D., Gadd~G.~E., Moricca~S., Sundqvist~B., Low Temp. Phys., 2005, \textbf{31}, No.~5, 429--444, \doi{10.1063/1.1925371}.
\bibitem{gugenberger1992glass} Gugenberger F., Heid R., Meingast C., Adelmann P., Braun M., W\"uhl H., Haluska M., Kuzmany~H., Phys.~Rev.~Lett., 1992, \textbf{69}, No.~26, 3774, \doi{10.1103/PhysRevLett.69.3774}.
\bibitem{reich2002elastic} Reich S., Thomsen C., Ordej\'on P., Phys. Rev. B, 2002, \textbf{65}, No.~15, 153407, \doi{10.1103/PhysRevB.65.153407}.
\bibitem{barron1982thermal} Barron T. H. K., Collins J. F., Smith T. W., White G. K., J. Phys. C: Solid State Phys., 1982, \textbf{15}, No.~20, 4311,\\ \doi{10.1088/0022-3719/15/20/016}.
\bibitem{miller2009negative} Miller W., Smith C. W., Mackenzie D. S., Evans K. E., J. Mater. Sci., 2009, \textbf{44}, No.~20, 5441--5451,\\ \doi{10.1007/s10853-009-3692-4}.
\bibitem{szewczyk2021heat} Szewczyk D., Gebbia J. F., Je\.zowski A., Krivchikov A. I., Guidi T., Cazorla C., Tamarit J. L., Sci. Rep., 2021, \textbf{11}, No.~1, 18640, \doi{10.1038/s41598-021-97973-2}.
\bibitem{miyazaki2021low} Miyazaki Y., Nakano M., Krivchikov~A.~I., Koroyuk O. A., Gebbia J. F., Cazorla C., Tamarit J. L., J.~Phys.~Chem.~Lett., 2021, \textbf{12}, No.~8, 2112--2117, \doi{10.1021/acs.jpclett.1c00289}.
\bibitem{strzhemechny2019heat} Strzhemechny M. A., Krivchikov~A.~I., Je\.zowski A., Low Temp. Phys., 2019, \textbf{45}, No.~12, 1290--1295,\\ \doi{10.1063/10.0000211}.
\bibitem{krause1980direct} Krause J. K., Swenson C. A., Phys. Rev. B, 1980, \textbf{21}, No.~6, 2533, \doi{10.1103/PhysRevB.21.2533}.

\end{thebibliography}

%

\ukrainianpart

\title{Пропорційна кореляція між теплоємністю та тепловим розширенням атомарних, молекулярних кристалів та вуглецевих наносистем}
\author{М.~С.~Барабашко\refaddr{label1}, О.~І.~Кривчіков\refaddr{label1, label2}, Р.~М.~Баснукаєва\refaddr{label1}, О.~О.~Королюк\refaddr{label1,label3}, 
А.~Єжовскі\refaddr{label2} }
\addresses{
\addr{label1} Фізико-технічний інститут низьких температур ім. Б.~І. Вєркіна Національної академії наук України,  Проспект Науки, 47, Харків, 61103, Україна, 1
\addr{label2} Інститут фізики низьких температур та структурних досліджень Польської академії наук, вул. Окольна, 2, 50-422, Вроцлав, Польща
\addr{label3} Міжнародний фізичний центр Доностія, Набережна Мануеля де Лардісабаля, 20018 Доностія - Сан Себаст'ян, Іспанія
}
%
%
%

\makeukrtitle

\begin{abstract}
\tolerance=3000%

Проаналізовано справедливість співвідношення між тепловим розширенням $\beta(T)$ і теплоємністю $C(T)$ атомарних і молекулярних кристалів, аморфних матеріалів зі структурним розладом, вуглецевих наноматеріалів (фуллерит C$_\textrm{60}$, системи компактованих джгутів одностінних вуглецевих нанотрубок). Розглянуто вплив внеску до коефіцієнту лінійного теплового розширення $\alpha_\textrm{Xe}(T)$  атомів Хе, адсорбованих механічно спресованими компактами з джгутів ОВНТ. Знайдена пропорційна кореляція $\alpha_\textrm{Xe}(T)/\alpha^* \approx  C_\textrm{Xe}(T)/R$ між внеском до коефіцієнту лінійного теплового розширення $\alpha_\textrm{Xe}(T)$ і нормалізованої на газову постійну теплоємністю $C_\textrm{Xe}(T)/R$ атомів Xe адсорбованих системою компактованих вуглецевих нанотрубок.
Пропорційна кореляція $(\beta/\beta^*) \approx  (C_\textrm{V}/R)$ з параметром перенормування $\beta^*$ для коефіцієнту об’ємного теплового розширення для кріокристалів було запропоновано аналогічно, як для коефіцієнту лінійного теплового розширення $\alpha_\textrm{Xe}(T)$. У випадку атомарних кристалів, таких як Xe, Ar, пропорційна кореляція $(\beta/\beta^*) \approx  (C_\textrm{V}/R)$ спостерігається у діапазоні температур від найнижчої експериментальної до температур, де $C_\textrm{V}/R \approx 2.3$ та не залежить від характеристик атомів цих кристалів. Зникнення кореляції спостерігається в області температур, де $2.3 < C_\textrm{V}/R < 3$ (класичної границі Дюлонга та Пті). Універсальна пропорційна кореляція також спостерігається для молекулярних кристалів з лінійною симетрією, таких як CO$_{2}$, CO та N$_{2}$O нижче значень теплоємності $C_\textrm{V}/R \approx 3 \div 3.5$, але відхилення від неї є більш пологим у порівнянні з атомарними кристалами. Ці факти вказують на те, що пропорційна кореляція пов’язана не тільки з трансляційними, але і з обертальними ступенями вільності молекули в кристалі. У випадку молекулярного кристалу С$_{60}$ з трансляційними, обертальними та внутрішніми ступенями вільності зазначена кореляція має місце до значень теплоємності $C_\textrm{V}/R \approx 7.5$. У сильно анізотропних системах, таких, як системи компактованих джгутів одностінних вуглецевих нанотрубок та системи компактованих вуглецевих нанотрубок із адсорбованими атомами Xe, ця універсальна залежність виконується в обмеженому інтервалі температур та не виконується при низьких температурах. Запропоновано якісне пояснення розглянутої кореляції.

\keywords теплоємність, теплове розширення, універсальне співвідношення, наноматеріали, низькорозмірні системи

\end{abstract}

\end{document}